\documentclass[journal,onecolumn]{IEEEtran}
\ifCLASSINFOpdf
\else
\fi

\usepackage{wrapfig}
\usepackage{graphics}
\usepackage[pdftex]{graphicx}
\usepackage{amssymb}
\usepackage{xspace}
\newtheorem{theorem}{Theorem}

\newtheorem{lemma}{Lemma}
\newtheorem{remark}{Remark}
\usepackage{fancyhdr}

\usepackage{algorithmic}

\usepackage{color, colortbl}
\newcommand{\vgms}{Variable-sized Generalized MS Code\xspace}

\newcommand{\onop}{online-optimal-rate\xspace}
\newcommand{\offop}{offline-optimal-rate\xspace}
\newcommand{\messagePacket}{message packet\xspace}
\newcommand{\messagePackets}{message packets\xspace}
\newcommand{\channelPacket}{channel packet\xspace}
\newcommand{\channelPackets}{channel packets\xspace}
\newcommand{\channelPacketsSinglePlural}{channel packet(s)\xspace}

\newcommand{\messageSizeSequence}{message size sequence\xspace}

\newcommand{\losslessDelay}{lossless-delay\xspace}
\newcommand{\worstCaseDelay}{worst-case-delay\xspace}
\newcommand{\regimeNoSpread}{Regime $1$\xspace}
\newcommand{\regimeTrivial}{Regime $2$\xspace}
\newcommand{\delay}{\tau}
\newcommand{\y}{Y}
\newcommand{\s}{S}
\newcommand{\x}{X}
\newcommand{\len}{t}
\newcommand{\burst}{b}
\newcommand{\identity}{\mathds{1}}

\newcommand{\maxTimeslot}{\lambda}

\newcommand{\messageSizeSequences}{message size sequences\xspace}
\usepackage[normalem]{ulem} 

\usepackage[utf8]{inputenc}

\usepackage{etoolbox}
\usepackage[dvipsnames]{xcolor}
\usepackage[normalem]{ulem} 

\newtoggle{showcomments}

\definecolor{color1}{HTML}{1b9e77}
\definecolor{color2}{HTML}{d95f02}
\definecolor{color3}{HTML}{7570b3}

\iftoggle{showcomments}{%
    \newcommand{\michael}[1]{%
        \textcolor{color1}{%
            \textbf{M says:} \textsf{%
            {#1}%
            }%
        }%
    }
    \newcommand{\mrem}[1]{%
        \textcolor{color1}{%
            \textbf{M removed:} {%
                {#1}%
            }%
        }%
    }%

        \newcommand{\rashmi}[1]{%
        \textcolor{color2}{%
            \textbf{R says:} \textsf{%
            {#1}%
            }%
        }%
    }
    \newcommand{\rrem}[1]{%
        \textcolor{color2}{%
            \textbf{R removed:} \sout{%
                {#1}%
            }%
        }%
    }%

}{%
    \newcommand{\michael}[1]{}
    
    \newcommand{\mrem}[1]{}
    \newcommand{\rashmi}[1]{}
    
    \newcommand{\rrem}[1]{}
}

\newcommand{\F}{\ensuremath{\mathbb F}}

\usepackage{subcaption}
\usepackage[T1]{fontenc}

\usepackage{ifthen}
\usepackage{cite}

\usepackage{dsfont}
\usepackage{url}
\usepackage{tikz}
\usepackage{textcomp}
\usepackage{lipsum}

\usepackage[cmex10]{amsmath}

\title{Online Versus Offline Rate \\in Streaming Codes for Variable-Size Messages
}

\author{Michael Rudow and K.V. Rashmi
\thanks{M. Rudow and K.V. Rashmi are with the Computer Science Department, Carnegie Mellon University, Pittsburgh,
PA, 15213 USA.}
\thanks{This paper was presented in part at~\cite{rudow2020Online} and published in full in~\cite{rudowonline2023Full}.}
}

\newcommand\copyrighttext{%
  \footnotesize \textcopyright  2023 IEEE.  Personal use of this material is permitted.  Permission from IEEE must be obtained for all other uses, in any current or future media, including reprinting/republishing this material for advertising or promotional purposes, creating new collective works, for resale or redistribution to servers or lists, or reuse of any copyrighted component of this work in other works. URL: \url{https://ieeexplore.ieee.org/document/10044142}
  DOI: 10.1109/TIT.2023.3244799}
\newcommand\copyrightnotice{%
\begin{tikzpicture}[remember picture,overlay]
\node[anchor=south,yshift=10pt] at (current page.south) {\fbox{\parbox{\dimexpr\textwidth-\fboxsep-\fboxrule\relax}{\copyrighttext}}};
\end{tikzpicture}%
}
\begin{document}
\title{Online Versus Offline Rate \\in Streaming Codes for Variable-Size Messages}

\maketitle
\copyrightnotice

\begin{abstract}

One pervasive challenge in providing a high quality-of-service for live communication is to recover lost packets in real-time.
Streaming codes are a class of erasure codes that are designed for such strict, low-latency streaming communication settings.
Motivated by applications that transmit messages whose sizes vary over time, such as live video streaming, this paper considers the setting of streaming codes under variable-size messages.
In practice, streaming codes operate in an ``online'' setting where the sizes of the future messages are unknown.
``Offline'' codes, in contrast, have access to the sizes of all messages, including future ones.
This paper introduces the first online rate-optimal streaming codes for communicating over a burst-only packet loss channel for two broad parameter regimes. These two online codes match the rates of optimal offline codes for the two settings despite the apparent advantage of the offline setting. This paper further establishes that online codes cannot attain the optimal rate for offline codes for all remaining parameter settings.

\end{abstract}


\pagestyle{plain}

\newcommand{\para}[1]{\smallskip\noindent\textbf{#1}~}
\section{Introduction}
\label{sec:intro}
Real-time communication with a high quality-of-service is critical for many pervasive streaming applications, including VoIP and videoconferencing. These live streaming applications rely on transmitting packets of information and contend with packet losses during transmission.
Although lost packets can be recovered via retransmission, this solution is often infeasible due to strict latency constraints~\cite{badr2017perfecting}. Therefore, real-time streaming applications often use forward error correction to provide robustness to packet losses. However, using traditional coding schemes to comply with the real-time delay constraint penalizes the rate.

\begin{figure}
\centering
\includegraphics[width=.5\textwidth]{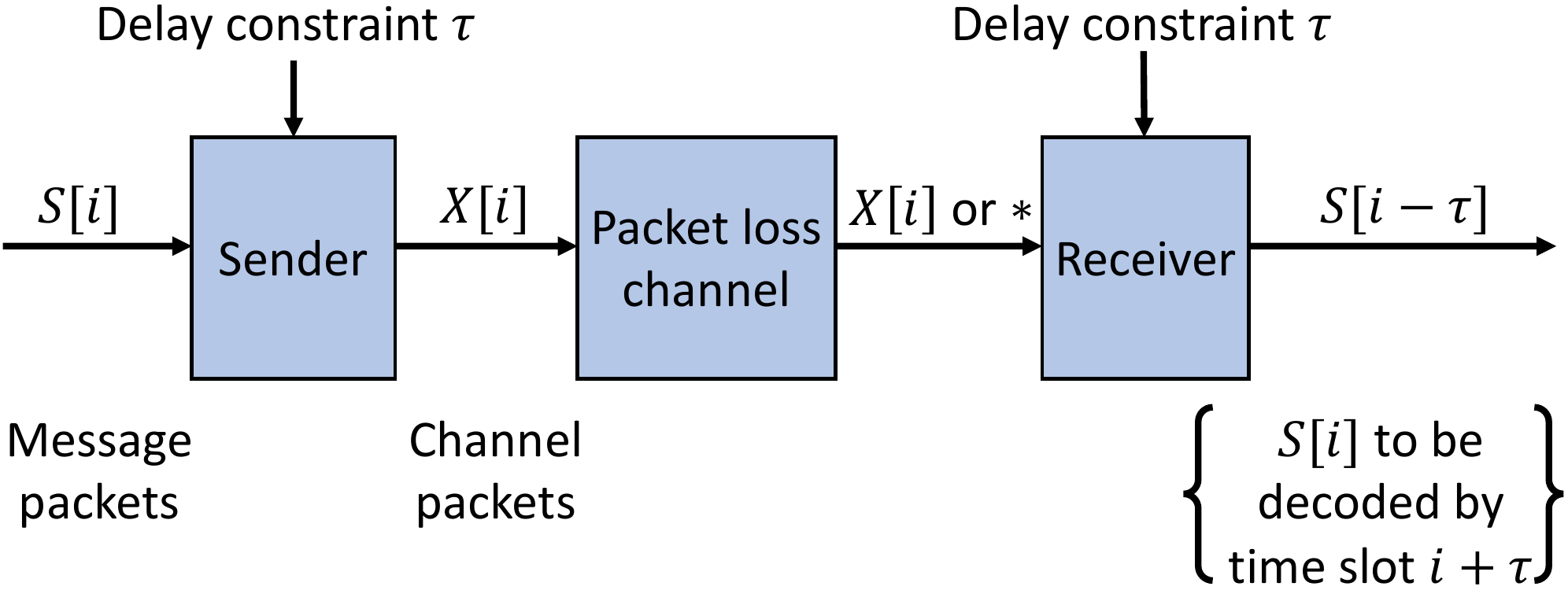}
    \caption{Overview of the streaming model. }
    \label{fig:model}
\end{figure}

Coding schemes explicitly designed for live streaming communication can attain significantly higher rates than traditional ones, such as maximal distance separable block codes. This improved performance was demonstrated in \cite{martinian2004burst}, where the authors proposed a new ``streaming model'' for real-time communication shown in Figure~\ref{fig:model}.
Under this streaming model, at each time slot $i$, a sender receives a ``\messagePacket'' $S[i]$ and transmits a ``\channelPacket'' $X[i]$ over a packet loss channel to a receiver. The message packet $S[i]$ is to be decoded at the receiver within delay $\delay$, i.e., by time slot $(i+\delay)$.
{The authors established an upper bound on the rate, and they introduced a rate-optimal construction for certain settings.
Later, a rate-optimal construction for all remaining settings was presented in~\cite{martinian2007delay}.}
{Numerous subsequent works have also studied variants of the streaming model} (discussed in Section~\ref{sec:modBack})~\cite{badr2017layered,fong2018optimalLong,krishnan2018rate,krishnan2019low,dudzicz2019explicit,krishnan2020rate,domanovitz2019explicit,badr2018multiplexed,fong2020optimalM,badr2011diversity,badr2015streamingM,haghifam2021streaming,adler2017burst,leong2012erasure,leong2013coding,badr2015streaming,Fong2020optimal,Badr2017FEC,li2011correcting,wei2013prioritized,su2021random,haghifam2018streaming,Fong2020optimal,rudow2018variable,rudow2020Online}.

The streaming model proposed in~\cite{martinian2004burst} and studied further in several subsequent works~\cite{badr2017layered,fong2018optimalLong,krishnan2018rate,krishnan2019low,dudzicz2019explicit,krishnan2020rate,domanovitz2019explicit,badr2018multiplexed,fong2020optimalM,badr2011diversity,badr2015streamingM,haghifam2021streaming,adler2017burst,leong2012erasure,leong2013coding,badr2015streaming,Fong2020optimal,Badr2017FEC,li2011correcting,wei2013prioritized,krishnan2019simple,su2021random,haghifam2018streaming,Fong2020optimal} considers a setting where all \messagePackets comprise some \textit{fixed} number of symbols.
However, many applications must send a stream of variable-size \messagePackets.
For example, video calls consist of compressed video frames of fluctuating sizes.
Consequently, a new streaming model incorporating \textit{variable-size} \messagePackets was introduced in \cite{rudow2018variable}.

The streaming model with variable-size \messagePackets differs from that of fixed-size \messagePackets in two key ways:
First, the sequence of sizes of \messagePackets affects the optimal rate. In fact, the variability in the sizes of \messagePackets \textit{negatively} impacts the optimal rate, which is never higher than that of the setting where \messagePackets have fixed sizes~\cite{rudow2018variable}. Second, while there are rate-optimal schemes that send each \messagePacket in the corresponding \channelPacket for the setting of fixed-size \messagePackets, spreading message symbols over multiple \channelPackets is advantageous in the setting of variable-size \messagePackets.
This is because sending a large \messagePacket within a single \channelPacket leads to many lost symbols when that \channelPacket is lost.
Spreading message symbols intelligently reduces the maximum number of message symbols lost in a burst\textemdash a lower bound on how much redundancy is needed.
In contrast, when all \messagePackets are the same size and are sent in the corresponding \channelPackets, all bursts drop the same number of message symbols.
As such, spreading message symbols over multiple \channelPackets does not offer an advantage.

When the transmission is lossless, sending message symbols over multiple \channelPackets increases the latency compared to sending each \messagePacket within the corresponding \channelPacket.
In~\cite{rudow2018variable}, the authors introduce a new delay constraint that captures the trade-off between the rate and the decoding delay in lossless transmission, called the \textit{\losslessDelay constraint}.
Specifically, when there are no losses, the receiver must decode each \messagePacket with a delay of $\delay_L$ time slots, {where $\delay_L$ is less than $\delay$}.\footnote{The \losslessDelay constraint was denoted as $\delay_G$ in~\cite{rudow2018variable}.
}

One key challenge in realizing the benefits of spreading is determining how to best spread message symbols over one or more \channelPacketsSinglePlural despite the fact that future \messagePackets' sizes are inherently variable and unknown.
For example, a large \messagePacket should be sent in the corresponding \channelPacket when the next several \messagePackets are even larger to reduce the variability in the sizes of \channelPackets.
In contrast, message symbols of a large \messagePacket should be spread over multiple \channelPackets when the subsequent several \messagePackets are small.
Thus, the optimal strategy for encoding depends on the sizes of future \messagePackets.
To capture this dependency introduced by the variability in the size of \messagePackets, the coding schemes can be classified into two classes: (a)``offline'' schemes and (b) ``online'' schemes.
Offline coding schemes have access to the \textit{sizes} of \messagePackets of \textit{future} time slots, whereas online schemes do not have access to such information.
Online constructions are of practical interest, as the sizes of future \messagePackets are typically unknown in live streaming applications.
By using future information, optimal offline constructions can always match, and potentially significantly exceed, the rate of online ones.
Therefore, a natural question is: ``can online coding schemes match the rate of offline coding schemes?''

\para{Main contributions.} In this paper, we {design the first rate-optimal online coding schemes for two classes of parameter settings.}
In ``\regimeNoSpread,'' $\burst$ and $\delay$ may take any values while $\delay_L=0$, necessitating that all constructions recover each \messagePacket immediately under lossless conditions\textemdash a useful property exhibited by existing rate-optimal constructions~\cite{martinian2004burst,martinian2007delay} for the streaming model where all \messagePackets have the same size.
This broad regime is well-suited for applications that require minimal latency during lossless conditions and can tolerate extra latency only during occasional losses.
Our rate-optimal construction is systematic, and it sends each \messagePacket in the corresponding \channelPacket.
During each time slot, $i$, we combine two new methodologies to alleviate the variability.
(a) We apply a greedy paradigm for delaying transmitting the parity symbols associated with $\s[i]$ until the time slot $(i+\delay)$.
(b) We define the number of parity symbols to be sent in $\x[i+\delay]$ while deferring defining the parity symbols themselves until the time slot $(i+\delay)$ to make use of the sizes of \messagePackets $\s[i+1],\ldots,\s[i+\delay-1]$.
The construction is rate-optimal, even for the offline setting. To prove the construction's optimality, we show that it cumulatively sends no more symbols by each time slot than any offline rate-optimal construction that satisfies the \worstCaseDelay and \losslessDelay constraints.
As such, the results show that non-systematic schemes provide no advantage.
In ``\regimeTrivial,'' $\tau_L = (\tau-b)$ and $b|\tau$, so $\delay_L$ has its maximum value.
Here, we show that a simple scheme that encodes each \messagePacket separately matches an upper bound on the rate.
Thus, the above results together show that online coding schemes can match the rate of optimal offline coding schemes for two broad parameter regimes even though knowledge about the sizes of future \messagePackets appears advantageous.
In addition, we demonstrate that online coding schemes necessarily have lower rates than optimal offline coding schemes for \textit{all} remaining parameter regimes.

The organization of the paper is as follows. We begin by introducing the model and background in Section~\ref{sec:modBack}. We then present online constructions {that match the optimal rate of offline constructions} for two parameter regimes in Section~\ref{sec:achieve}.
Next, we show that the rate of optimal online schemes cannot match that of offline schemes for all remaining settings in Section~\ref{sec:gap}.
Finally, we end with a discussion on conclusions and future directions in Section~\ref{sec:conclusion}.

\section{Background, System model, and related work}
\label{sec:modBack}
We begin this section by discussing the background on streaming codes that led to the model considered in this work. We then present the model in detail, as well as the notation used throughout this paper. Finally, we discuss related work on streaming codes.

\subsection{Background}

Martinian and Sundberg proposed the streaming model in~\cite{martinian2004burst}.
It captures the setting of real-time communication of a sequence of \messagePackets of a fixed size over a burst-only packet loss channel.
At each time slot $i$, a sender receives a \textit{\messagePacket}, $S[i]$, comprising $k$ symbols drawn uniformly at random from a finite field $\F_q$.
The sender then transmits to a receiver a \textit{channel packet}, $X[i]$, consisting of $n$ symbols from $\F_q$ over a burst-only channel.
Due to real-time latency constraints, the receiver must decode $S[i]$ within a delay of $\delay$ time slots (that is, using the channel packets received by time slot $(i+\delay)$).
The lossy channel is denoted $C(\burst,\delay)$ and introduces bursts of length at most $\burst$ followed by guardspaces of length at least $\delay$.
The authors showed an upper bound on the rate of streaming codes of
$\frac{\delay}{\delay+\burst}$ and introduced a class of code constructions applicable to the streaming model, called ``streaming codes,'' meeting this bound for some settings of $\delay$ and $\burst$.
Later, a construction proposed in~\cite{martinian2007delay} met this bound{, showing that $\frac{\delay}{\delay+\burst}$ is the capacity for the remaining settings of $\delay$ and $\burst$}.

In applications such as video communication, the sizes of messages fluctuate considerably. Consequently, in~\cite{rudow2018variable} a streaming model was introduced that incorporates variable-size messages.
The authors showed that $\frac{\delay}{\delay+\burst}$ remains an upper bound on the rate.
The authors also present a streaming code for this new setting, and via an
empirical evaluation, show that the construction attains a rate of approximately $89.5\%$ of the upper bound on rate of $\frac{\delay}{\delay+\burst}$ for the settings considered in the empirical evaluation.
The authors also bounded the gap between the construction and $\frac{\delay}{\delay+\burst}$ when the sizes of \messagePackets are drawn independently from a distribution.
The smaller the variance of the distribution, the smaller the gap.
However, the gap is nontrivial, and the sizes of \messagePackets for real-time streaming applications are typically not independent.

\subsection{System model}
\label{sec:sytemModel}
We consider the streaming model from \cite{rudow2018variable}, which considers variable-size \messagePackets (with a few minor changes in {how time slots are indexed}).
During each time slot $i$ the \messagePacket, $S[i]$, comprises $k_i \in \{0,\ldots,m\}$ symbols for a natural number $m$ representing the maximum possible size of a \messagePacket.
The sender transmits a \channelPacket, $\x[i]$, comprising $n_i$ symbols.
The receiver obtains
\begin{equation*}
\y[i] =
     \begin{cases}
    \x[i] & \text{if $\x[i]$ is received}\\
    * & \text{if $\x[i]$ is lost}.
\end{cases}
\end{equation*}
Transmission occurs over a $C(\burst,\delay)$ channel.
Each channel packet, $X[i]$, depends only on the symbols of previous \messagePackets (i.e. $S[0],\ldots,S[i]$).
Similar to the model of fixed-size \messagePackets, each $\s[i]$ must be decoded by time slot $(i+\delay)${; this requirement is called the \textit{\worstCaseDelay constraint}}.

Recall from Section~\ref{sec:intro} that under the setting of variable-size \messagePackets,
spreading message symbols over multiple \channelPackets can be advantageous.
As such, there is an inherent tradeoff between the rate of a code and the decoding delay under lossless transmission (i.e., the number of time slots needed to decode a \messagePacket when all channel packets are received).
A new delay constraint capturing this trade-off, called the \textit{\losslessDelay constraint}, was introduced in~\cite{rudow2018variable}. When there are no losses, the receiver must decode each \messagePacket $S[i]$ within a delay of $\delay_L$ (< $\delay$) time slots.
{The \losslessDelay constraint is relevant for applications that can infrequently tolerate a delay of $\delay$ in the worst case but require faster decoding for most \messagePackets.}

The valid value ranges for the parameters $b,\delay,$ and $\delay_L$ are $1 \le b \le \delay$ and $0 \le \delay_L \le (\delay-b)$. A maximum burst length of $0$ is omitted because coding is unnecessary for lossless transmission. Furthermore, reliable transmission is impossible when $b$ exceeds $\delay$, since $S[i]$ cannot be decoded by time slot $(i+\delay)$ when $X[i],\ldots,X[i+\delay]$ are all lost in a burst. Intrinsically, $\delay_L$ cannot be negative, and $S[i]$ is decoded by time slot $(i+\delay-b)$ if there are no losses, since a burst can drop $X[i+\delay-b+1],\ldots,X[i+\delay]$. Since $b >0$, this means that $\delay_L$ is without loss of generality strictly less than $\delay$.

{In the setting where \messagePackets all have size $k$ and \channelPackets all have size $n$~\cite{martinian2004burst}, the rate is $\frac{k}{n}$.
However, the setting of varying sizes of \messagePackets and \channelPackets, necessitates a new definition of rate.
The rate is defined~\cite{rudow2018variable} for a finite stream of $(\len+1)$ \messagePackets for an arbitrary natural number $\len$ as the number of message symbols divided by the number of transmitted symbols:}
\begin{equation*}
    R_\len= \frac{\sum_{i=0}^\len k_i}{\sum_{i=0}^\len n_i}
\end{equation*}

Recall that the rate is at most $\frac{\delay}{\delay+\burst}$.
However, depending on the sizes of the \messagePackets, the upper bound can be loose.

Constructions that during the time slot $i \in \{0,\ldots,\len\}$ can access all future \messagePackets' sizes (i.e., $k_{i+1},\ldots,k_\len$) are called ``offline.''
Offline schemes have access to the \textit{sizes} but not the \textit{symbols} of the future \messagePackets.
In contrast, code constructions that do not know the sizes of the future \messagePackets are dubbed ``online.'' Specifically, during time slot $i$, $(k_{i+1},\ldots,k_\len)$ are unknown for an online construction.
We distinguish between the feasible rates for offline and online coding schemes.
The best possible rate for offline coding schemes is called the ``\offop'' and for online coding schemes is called the ``\onop.''

Encoding during time slot $i$ is defined as
\begin{equation}
\label{eq:encode}
\x[i] = Enc\left(\s[0],\ldots,\s[i]\right).
\end{equation}
To distinguish between online and offline decoding, we use the following quantity to denote the last time slot for which the size of \messagePackets is available {to the receiver}
\begin{equation*}
\maxTimeslot_i = \begin{cases}
    \len & \text{if offline}\\
    \arg \max_{l \in \{i,\ldots,i+\delay\}} \identity\left[\y[l] == \x[l]\right]   & \text{if online}.
\end{cases}
\end{equation*}
The decoding for \messagePacket $\s[i]$ is then defined for two scenarios.
First, in a lossless transmission, $\s[i]$ is decoded using (a) the previously decoded \messagePackets, (b) the $(\delay_L+1)$ \channelPackets received within \losslessDelay, and (c) the sizes of the first $(i+\delay_L+1)$ \messagePackets as follows:
\begin{equation}
\label{eq:decodeLossless}
\s[i] = Dec^{(L)}\big(\s[0],\ldots,\s[i-1],\x[i],\ldots,\x[i+\delay_L],k_{0},\ldots,k_{i+\delay_L}\big).
\end{equation}
Second, when losses occur, $\s[i]$ is decoded using (a) the previously decoded \messagePackets, (b) all received \channelPackets among the $(\delay+1)$ sent within the \worstCaseDelay, and (c) the sizes of the first $(\maxTimeslot_i + 1)$ \messagePackets as follows:
\begin{equation}
\label{eq:decode}
\s[i] = Dec\big(\s[0],\ldots,\s[i-1],\y[i],\ldots,\y[i+\delay],k_{0},\ldots,k_{\maxTimeslot_{i+\delay}}\big).
\end{equation}
To ensure that the receiver knows the sizes of \messagePackets, a small header containing $k_{i-\burst},\ldots,k_i$ is added to $\x[i]$.\footnote{{In the edge conditions, $(i-\delay)$ is set to $0$ for $i<\delay$ and $(i+\delay)$ is set to $\len$ for $(i-\delay)>(\len-\delay)$.}}
Finally, we note that our work's constructions do not need as much memory as is acceptable under the model. During any time slot, $i$, the sizes and symbols of \messagePackets and \channelPackets from before time slot $(i-\delay)$ are not used.

The capacity is defined for any given \messageSizeSequence, $k_0,\ldots,k_\len,$ as the highest rate that can be attained while satisfying Equations~\ref{eq:encode},~\ref{eq:decodeLossless}, and~\ref{eq:decode}.

This paper uses the following notation.
The term $[n]$ denotes $\{0,\ldots,n\}$.
All vectors are row vectors. A vector $V$ has length $v$ and is indexed as $V = (V_0,\ldots,V_{v-1})$.
For $I=\{i_0,\ldots,i_{l}\} \subseteq [v-1]$ where $i_j < i_{j'}$ for $j<j' \in [l]$, $V_I = (V_{i_0},\ldots,V_{i_{l}})$.
Let $A$ be an $n \times n$ matrix, and $I\subseteq \{0,\ldots,n-1\}$. Then $A_{I}$ is $A$ restricted to the columns in $I$.
This work refers to $k_0,\ldots,k_\len$ as the ``\messageSizeSequence.''

This work uses the following conventions.
The sizes of the final $\delay$ \messagePackets are each $0$, and $\len$ is at least $\delay$.
Thus, the coding schemes can encode the final \messagePacket of non-zero size using $\delay$ extra channel packets.
{To satisfy this restriction, one can append $\delay$ \messagePackets of size $0$ to the stream of messages, which will not change the optimal rate.}

For $i \in \{1-\burst,\ldots,-1\} \cup \{\len+1, \ldots,\len+\burst+1\}$, $k_i$ is defined as $0$. For $i \in \{1-\burst,\ldots,-1\}$, a burst loss of $X[i],\ldots,X[i+\burst-1]$ denotes a burst loss of $X[0],\ldots,X[i+\burst-1]$.
Similarly, for $i \in \{\len-\burst+2,\ldots,\len\}$ a burst loss of $X[i],\ldots,X[i+\burst-1]$ denotes a burst loss of $X[i],\ldots,X[\len]$.

\subsection{Other related works}
{Numerous existing works have examined different variations of the streaming model introduced by Martinian and Sundberg in~\cite{martinian2004burst}.
These streaming models involve fixing the sizes of \messagePackets and \channelPackets in advance.
Badr et al.~\cite{badr2017layered} introduced a new streaming model with fixed-size \messagePackets and \channelPackets in which every sliding window of $w$ \channelPackets can include (a) a burst of length $\burst$ or (b) up to $a$ arbitrary losses.
The authors also showed an upper bound on the rate under this sliding window model of loss.
Several later works~\cite{krishnan2019low,fong2018optimalLong,krishnan2018rate,dudzicz2019explicit,domanovitz2019explicit,krishnan2020rate} designed streaming codes that matched this upper bound on the rate.
Two previous works~\cite{badr2018multiplexed,fong2020optimalM} studied the setting of multiplexing two streams of \messagePackets with different delay constraints.
A few works~\cite{badr2011diversity,badr2015streamingM,Badr2017FEC} have considered streaming codes where there are two different decoding delay constraints based on two different types of packet loss.
In~\cite{haghifam2021streaming}, the authors studied the setting where all or some symbols of \messagePackets are recovered for short or long bursts, respectively.
Badr et al. investigated~\cite{badr2015streaming} streaming codes that recover only some \messagePackets within the delay constraint, depending on the loss patterns.
Another work~\cite{adler2017burst} studied streaming codes in terms of the average decoding delay rather than the maximum delay.
In~\cite{su2021random}, the authors evaluate the trade-off between memory, decoding delay, and decoding probability for random linear streaming codes with i.i.d. losses.
Several works~\cite{badr2017layered,leong2012erasure,leong2013coding} studied models of streaming codes where multiple \channelPackets are sent during each time slot.
In~\cite{li2011correcting}, the authors presented streaming codes to recover multiple bursts within $(\delay+1)$ \channelPackets.
Another work~\cite{wei2013prioritized} considered unequal error protection for a streaming model with high and low priority messages of two different fixed sizes when the sequence of the priorities of the messages is periodic.
Several recent works~\cite{Fong2020optimal,krishnan2021high,domanovitz2020streaming} have applied streaming codes to multi-node relay networks.
Future work could compare online and offline constructions for these variants of the streaming model after incorporating \messagePackets of varying sizes. }

\section{Online Code Constructions with Optimal Rate}
\label{sec:achieve}

In this section, we present the first rate-optimal online {streaming codes}, as well as show that they match the \offop, for two broad parameter regimes{: \textit{\regimeNoSpread:} ($\tau_L=0$ and any $\burst$ and $\delay$}) and \textit{\regimeTrivial:} ($\tau_L = (\tau-b)$ and $b|\tau$).

To begin, we consider \regimeNoSpread (i.e., $\delay_L = 0$ and any $\burst$ and $\delay$). In this regime, the \losslessDelay constraint, $\delay_L = 0$, eliminates the choice of distributing symbols corresponding to a \messagePacket over multiple channel packets. We introduce a systematic construction that sends each \messagePacket within the corresponding \channelPacket.
The construction employs an online greedy paradigm for sending parity symbols. The approach involves (a) identifying during time slot $i$ how many parity symbols will be sent during time slot $(i+\delay)$ (i.e., in advance $\delay$ time slots), and (b) defining the parity symbols only during time slot $(i+\delay)$ based on the sizes of $\s[i+1],\ldots,\s[i+\delay-1]$. To show that the construction is rate-optimal, we demonstrate via induction that the cumulative number of symbols sent by each time slot $i \in [\len]$ is no more than that which is sent under an arbitrary offline construction.

We next present the rate-optimal online coding scheme for any $(\tau,b)$ under \regimeNoSpread.
The scheme builds on top of the Generalized Maximally Short Codes presented in~\cite{badr2017layered} in such a way so as to {mitigate the adverse effects of the} variability of the \messageSizeSequence. We call the proposed scheme the \textbf{($\tau,b$)-\vgms}.
The construction is suitable for any field of size at least $2\tau m$.
{We first provide a high-level description, then present a toy example, and finally present the details of the code construction.}

{\textbf{Encoding (high level description).} During time slot $i$, each \messagePacket $S[i]$ is partitioned into two pieces: $S[i]=(U[i],V[i])$. The channel packet $X[i]=(S[i],P[i])$ is then sent, where $P[i]$ comprises parity symbols. The parity symbols are defined as $P[i]=(U[i-\tau]+P'[i])$ where $P'[i]$ consists of carefully designed linear combinations of the symbols of $(V[i-\tau],\ldots,V[i-1])$.
{The linear equations are defined so that that for all $i \in [\len-\delay-\burst+1],$ $P'[i+\burst],\ldots,P'[i+\delay-1], V[0],\ldots,V[i-1]$ are sufficient to decode $V[i],\ldots,V[i+\burst-1]$, as will be fully explained in the detailed description}.}\footnote{For $i<\tau$, $P[i]$ is empty.}

We set $V[i]$ to contain as many symbols of $S[i]$ as possible while meeting the following requirement. For any $j \in \{i-\burst+1,\ldots,i\}$ and burst loss of $X[j],\ldots,X[j+b-1]$, the sum of the sizes of $V[j],\ldots,V[i]$ is at most the number of parity symbols in $X[j+b],\ldots,X[j+\tau-1]$ (i.e., the sum of the sizes of $P[j+b],\ldots,P[j+\tau-1]$).
The remaining symbols of $S[i]$ are allocated to $U[i]$.
The size of $P[i]$ is set to equal that of $U[i-\tau]$.

\textbf{Decoding (high level description).} A burst loss of $X[i],\ldots,X[i+b-1]$ is recovered in two steps. First, for $j \in \{i+b,\ldots,i+\tau-1\}$, $U[j-\tau]$ is subtracted
from $P[j]$ to obtain $P'[j]$. Then $P'[i+b],\ldots,P'[i+\tau-1]$ are used to recover $V[i],\ldots,V[i+b-1]$ during the time slot $(i+\tau-1)$.
Recovery is possible because (a) $P'[i+b],\ldots,P'[i+\tau-1]$ contain at least as many symbols as $V[i],\ldots,V[i+\burst-1]$ by definition, and (b) the linear equations used to define $P'[i+b],\ldots,P'[i+\tau-1]$ are chosen to be linearly independent.
Second, during time slot $j \in \{i+\tau,\ldots,i+\tau+b-1\}$, $V[j-\tau],\ldots,V[j-1]$ are used to compute $P'[j]$. Subtracting $P'[j]$ from $P[j]$ yields $U[j-\delay]$.

\begin{figure}
\centering
\includegraphics[angle=0,width=.5 \textwidth, trim = {0 0 2cm 7.5cm},clip]{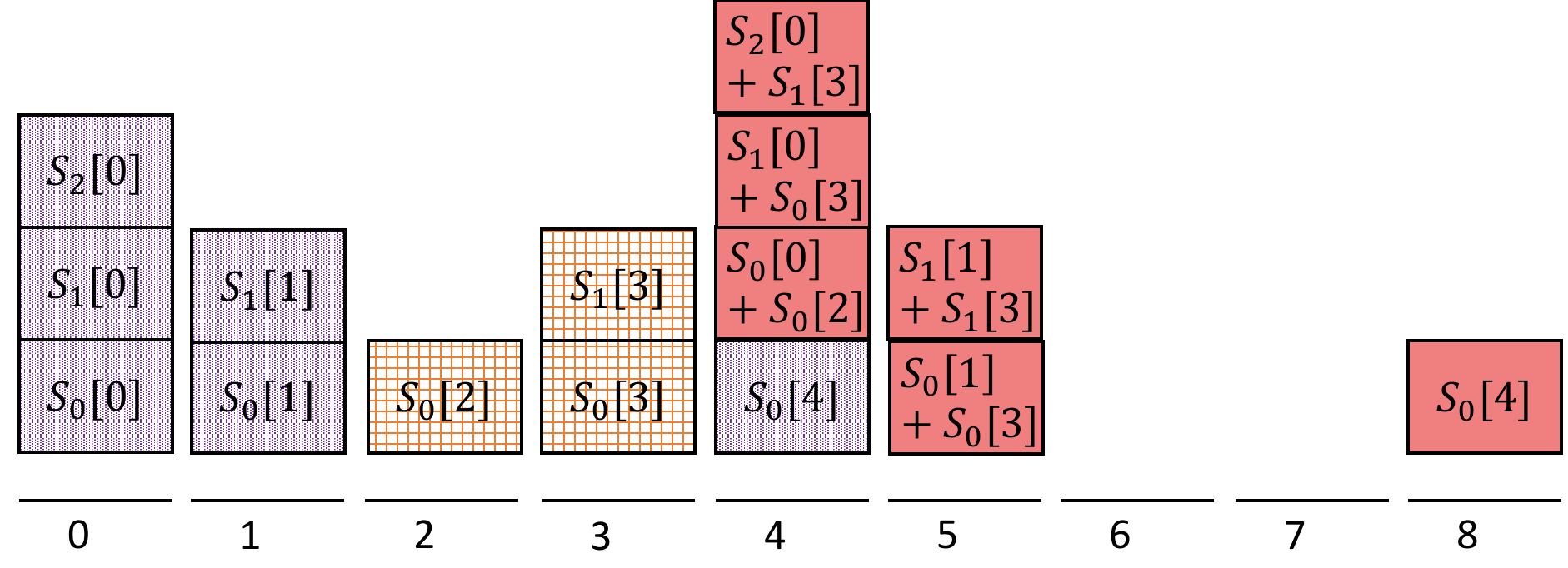}

\caption{{A toy example of the $(\tau = 4, \burst = 2)$-\vgms. Each \messagePacket, $S[i] = (U[i],V[i])$, is transmitted in the corresponding channel packet, $X[i]$, along with parity symbols, $P[i]$, (when applicable). White boxes with purple dots represent symbols of $U[i]$, white boxes with an orange grid represent symbols of $V[i]$, and solid red boxes represent symbols of $P[i]$. The numbers under the lines at the bottom indicate the time slots.}}
 \vspace{-10pt}
\label{fig:toyVGMS}
\end{figure}

\textbf{Code construction (toy example).} {We now present a toy example of $(\tau = 4, \burst =2)-$\vgms for \messageSizeSequence $k_0 = 3, k_1 = 2, k_2 = 1, k_3=2,k_4=1,$ and $k_5 =\ldots = k_8 = 0$, shown in Figure~\ref{fig:toyVGMS}.
For $i \in [4]$, $S[i]$ is sent in $X[i]$. This satisfies the \losslessDelay constraint. For $i \in \{0,1,4\}$, $U[i]$ is defined to equal $S[i]$, and $V[i]$ is defined to be empty (i.e., of size $0$).
For $i \in \{2,3\}$, $V[i]$ is set as $S[i]$, and $U[i]$ is defined to be empty.
{Let $P'[4] = (S_0[2],S_0[3],S_1[3])$ and $P'[5] = (S_0[3],S_1[3])$. Next, $P[4] = \left(S[0] + P'[4]\right)$ is transmitted in $X[4]$, and $P[5] = (S[1] + P'[5])$ is sent in $X[5]$. Finally, $P_0[8] = S_0[4]$ is transmitted in $X[8]$.
The \losslessDelay constraint is met, since each \messagePacket is sent within the corresponding \channelPacket. If any symbols of $V[2]$ and or $V[3]$ are lost, they are recovered using $P[4]$ and $P[5]$ respectively.
Any lost symbols of $U[0],U[1],$ and $U[4]$ are each decoded with delay exactly $4$ using $P[4],P[5],$ and $P[8]$ respectively (and subtracting $P'[4]$ and $P'[5]$ from $P[4]$ and $P[5]$ respectively).}
Therefore, the \worstCaseDelay constraint is satisfied.}

{Before presenting the detailed description, we introduce some notation. For $Z \in \{\s,\x,U,V,P,P'\}$ and any $i \le j \in [\len]$, $Z[i]$ is a vector of length $z[i]$, and $Z[i:j] = \left(Z[i],\ldots,Z[j]\right)$.}

\textbf{Code construction (detailed description).}
During each time slot $i$, the channel packet $X[i] = (S[i],P[i])$ is sent. The scheme is formally described in three parts: initialization, partitioning $S[i]$ into $(U[i],V[i])$, and defining $P[i]$.

\textit{Initialization:} For $i \in [b-1]$, we set $U[i] = S[i]$ and $v[i]=0$. For $i \in [\tau-1]$ we set $p[i] = 0$.
Let $A$ be a $\tau m \times \tau m$ Cauchy matrix, where $m$ was defined in Section~\ref{sec:sytemModel} as an upper bound on the sizes of \messagePackets.

\textit{Partitioning $S[i]$:}
For any $i \ge b$, we partition $S[i]$ into $S[i] = (U[i],V[i])$ as follows.\footnote{{Recall that partitioning was defined for $i<\burst$ in initialization.}}
We define an auxiliary variable $z_i$ encapsulating the minimum number of parity symbols available for recovering $S[i]$ when $X[i]$ is dropped in a burst:
\begin{equation}
\label{eq:defZ}
z_i = \min_{j\in \{i-b+1,\ldots,i\}} \sum_{l = j+b}^{i+\tau-1} p[l] - \sum_{l=j}^{i-1} k_l.
\end{equation}
The first $\min(k_i,z_i)$ symbols of $S[i]$ are set to $V[i]$:
\begin{equation}
\label{eq:defV}
V[i] = \left(S_0[i],\ldots,S_{\min(k_i,z_i)-1}[i]\right)
\end{equation}
The remaining symbols of $S[i]$ are set to $U[i]$:
\begin{equation}
\label{eq:defU}
U[i] = \left(S_{\min(k_i,z_i)}[i],\ldots,S_{k_i-1}[i]\right).
\end{equation}
Finally,
\begin{equation}
\label{eq:numParity}
p[i+\tau] = u[i] = k_i - \min(k_i, z_i) = k_i - v[i]
\end{equation}
parity symbols are assigned to be sent in the channel packet $X[i+\tau]$, although the actual symbols of $P[i+\tau]$ have not yet been identified.
{The size of $p[i+\tau]$ is never greater than $k_i$ (that is, the maximum possible size of $u[i]$), therefore $p[i+\tau]$ is at most $m$. }

\begin{figure*}
\centering
\includegraphics[angle=0,width=\textwidth]{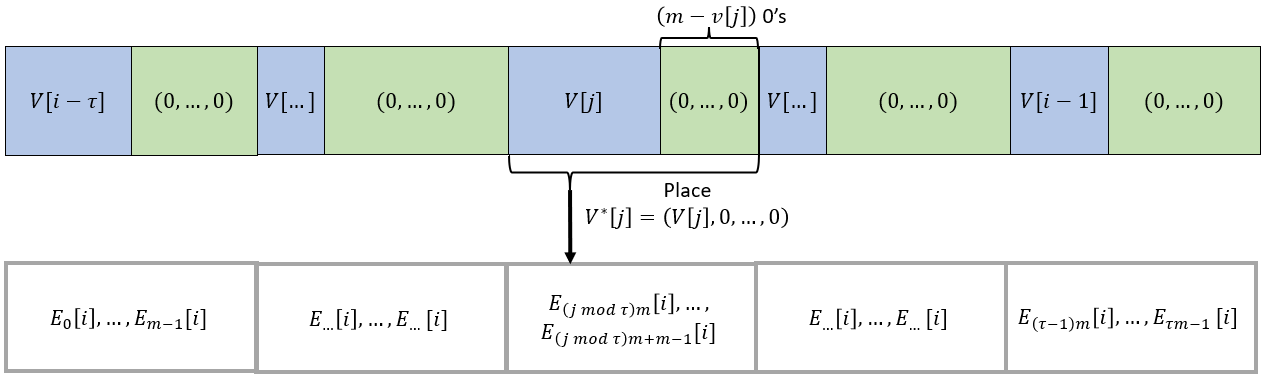}

\caption{Illustration for defining $E[i]$, for time slot $i \in [\len]$, by placing $V^*[j] = (V[j],0,\ldots,0)$, for $j \in \{i-\delay,\ldots,i-1\}$, into $m$ consecutive positions of $E[i]$ starting with position $(j \mod \delay)m$.
}
 \vspace{-10pt}
\label{fig:defineE}
\end{figure*}

\textit{Defining $P[i]$:}
During time slot $(i \ge \tau)$, we set
\begin{equation}
\label{eq:defParity}
P[i] = (U[i-\tau] + P'[i])
\end{equation}
where the symbols of $P'[i]$ are linear combinations of the symbols of $V[i-\tau],\ldots,V[i-1]$.\footnote{{Recall that $p[i]$ was defined during initialization for $i < \tau$.}}
{The linear combinations are chosen from a Cauchy matrix, as described below.}
Let $V^*[j]$ be the length $m$ vector obtained by appending ($m-v[j]$) $0$'s to $V[j]$ for $j \in \{i-\tau,\ldots,i-1\}$.
We define a vector of length $\tau m$, $E[i]$, by placing $V^*[j]$, for $j \in \{i-\tau,\ldots,i-1\}$,
into $m$ consecutive positions of $E[i]$ starting with position $(j \mod \tau)m${, as is detailed in Figure~\ref{fig:defineE}}.\footnote{For each $l \in \{i,\ldots,i+\delay-1\},$ $V^*[i]$ appears in the same positions of $E[l]$ as in $E[i]$. }
We use the Cauchy matrix A to define
\begin{equation}
\label{eq:parityCauchy}
P'[i] = E[i]A_{\{(i \mod \tau)m,\ldots, (i\mod \tau)m+p[i]-1\}}.
\end{equation}
The field size requirement is dictated by the Cauchy matrix and is at most $2\tau m$.

In Theorem~\ref{thm:osc_guarentee} below, we verify that the \vgms meets the requirements of the model.

 \begin{theorem}
For any parameters $(\tau,b)$ and \messageSizeSequence $k_0,\ldots,k_\len$, the ($\tau,b$)-\vgms satisfies the \losslessDelay and \worstCaseDelay constraints over any $C(b,\tau)$ channel.
 \label{thm:osc_guarentee}
 \end{theorem}
\begin{IEEEproof}
The \losslessDelay constraint is satisfied for $i \in [\len]$ by sending $X[i] = (S[i],P[i])$.

We prove that the \worstCaseDelay constraint is satisfied by showing for any $i \in [\len-\tau]$ that each of $S[i],\ldots,S[i+b-1]$ are recovered within delay $\tau$ when $X[i],\ldots,X[i+b-1]$ are lost.\footnote{Each \messagePacket $S[i]$ for $i>(\len-\tau)$ is of size $0$ and is known by the receiver due to the termination of the \messageSizeSequence.}
First, we show that $V[i],\ldots,V[i+b-1]$ are recovered by time slot $(i+\tau-1)$.
Second, we show that $U[i],\ldots,U[i+b-1]$ are recovered by time slots $(i+\delay),\ldots,(i+\delay+\burst-1)$, respectively.

First, for $j \in \{i+b,\ldots,i+\tau-1\}$ subtracting $U[j-\tau]$ from $P[j]$ yields $P'[j]$ (by Equation~\ref{eq:defParity}).
Combining Equations~\ref{eq:defV},~\ref{eq:defU},~\ref{eq:numParity}, and~\ref{eq:defParity} shows that the total number of symbols in $P'[i+b],\ldots,P'[i+\tau-1]$ is at least as many as $V[i],\ldots,V[i+b-1]$:
\begin{align*}
\sum_{j=i+b}^{i+\tau+b-1}p'[j]  &\ge \sum_{j=i}^{i+b-1} k_j \\
\sum_{j=i+b}^{i+\tau-1}p'[j]   + \sum_{j=i+\tau}^{i+\tau+b-1}p'[j]    &\ge  \sum_{j=i}^{i+b-1} v[j] +  \sum_{j=i}^{i+b-1} u[j]\\
\sum_{j=i+b}^{i+\tau-1}p'[j]  &\ge \sum_{j=i}^{i+b-1} v[j].
\end{align*}

{Next, we show that $P'[i+b],\ldots,P'[i+\tau-1]$ suffices to decode $V[i],\ldots,V[i+\burst-1]$. For $j \in \{i+\burst,\ldots,i+\delay-1\}$, recall from Equation~\ref{eq:parityCauchy} and Figure~\ref{fig:defineE} that $P'[j]$ is the product of distinct columns of $A$ with a vector consisting of (a) for $l \in \{i,\ldots,i+\burst-1\}$, $V[l]$ in positions $(j \mod \tau)m, \ldots,\left((j \mod \tau)m+v[l]-1\right)$, (b) for $l \in \{i,\ldots,i+\burst-1\}$, zeros in positions $\left((j \mod \tau)m+v[l]\right), \ldots, \left((j \mod \tau+1)m-1\right)$, and (c) a combination of symbols of $V[j-\delay],\ldots,V[i-1],V[i+\burst],\ldots,V[j-1]$ and zero padding in the remaining positions.
For $l \in \{i+\burst,\ldots,i+\delay-1\}$, let $E'[l]$ be defined by first setting it equal to $E[l]$ and second replacing the symbols corresponding to $V[i],\ldots,V[i+\burst-1]$ with $0$'s. We note that for $r \in \{i+\burst,\ldots,i+\delay-1\}$, the receiver can compute $E'[r]$ during time slot $(i+\delay-1)$.
Let $P^*[r]$ correspond to $\left(P'[r] - E'[r]A\right)$.
Then for some $l_0,\ldots,l_{\burst-1}$ which is a permutation of $i,\ldots,(i+\burst-1)$,}
\begin{equation*}
\begin{bmatrix}
P^*[i+b]^T \\
\vdots \\
P^*[i+\delay-1]^T
\end{bmatrix} =
\begin{bmatrix}
V[l_0]^T \\
\vdots \\
V[l_{\burst-1}]^T
\end{bmatrix}^TA'
\end{equation*}
{where $T$ denotes transpose, and $A'$ is a submatrix of $A$  with $\left(\sum_{j=i}^{i+\burst-1} v[j] \right)$ rows and at least $\left(\sum_{j=i}^{i+\burst-1} v[j] \right)$ columns. As such, $A'$ is Cauchy and thus has full rank. Hence, $P'[i+b],\ldots,P'[i+\tau-1]$ suffices to decode $V[i],\ldots,V[i+b-1].$}

Second, for $j \in \{i,\ldots,i+b-1\}$, $V[j],\ldots,V[j+\tau-1]$ are used to compute
\begin{equation*}
P'[j+\tau]=E[j+\tau]A_{\{(j \mod \tau)m,\ldots,(j\mod \tau)m+ p[j+\tau]-1\}}.
\end{equation*}
During time slot $(j+\tau)$, $U[j] = (P[j+\tau]-P'[j+\tau])$ is then decoded.\footnote{In the edge case where $i>(\len-\tau)$, $S[i]$ is known by the decoder to have size $0$ and this step is not needed.}

\end{IEEEproof}

The following lemma essentially shows that all parity symbols sent in any channel packet under the $(\tau,b)-$\vgms are needed to satisfy the \worstCaseDelay constraint.
{This property is later used to prove that the $(\tau,b)-$\vgms is rate-optimal in Theorem~\ref{thm:osc_opt}.}

\begin{lemma}
\label{lem:exactRec}
Consider any parameters $(\tau,b)$, \messageSizeSequence $k_0,\ldots,k_\len$, and the ($\tau,b$)-\vgms. For all $i \ge \tau$ where $p[i]>0$, $\exists j \in \{i-\tau-b+1,\ldots,i-\tau\}$ such that $\sum_{l=j}^{i-\tau} k_l=\sum_{l=j+b}^{i} p[l] $.
\end{lemma}

\begin{IEEEproof}
{For $i \in \{\tau,\ldots,\tau+b-1\}$, consider $j=0$.
Then
\begin{equation*}
\sum_{l=j}^{i-\tau} k_l = \sum_{l=0}^{i-\tau} u[l] = \sum_{l=\tau}^i p[l] = \sum_{l=\burst}^i p[l]
\end{equation*}
due to Equation~\ref{eq:numParity} as well as the initialization defining (a) $p[0],\ldots,p[\tau-1]$ to each be 0, and (b) $u[0],\ldots,u[\burst-1]$ to be $k_0,\ldots,k_{\burst-1}$ respectively.} 

For $(i \ge \tau+b)$, if $(p[i]=u[i-\delay]>0)$ then $(v[i-\delay]<k_{i-\tau})$. {By Equations~\ref{eq:defZ} and~\ref{eq:defV} and the fact that $(v[i-\delay]<k_{i-\tau})$ there is some $j \in \{i-\tau-b+1,\ldots,i-\tau\}$ for which for $i' = (i-\delay)$}
\begin{align*}
 v[i'] &= \sum_{l = j+b}^{i'+\delay-1} p[l] - \sum_{l=j}^{i'-1} k_l\\
 v[i-\delay] &= \sum_{l = j+b}^{i-1} p[l] - \sum_{l=j}^{i-\tau-1} k_l\\   v[i-\delay] + u[i-\delay] + \sum_{l=j}^{i-\tau-1} k_l &= p[i] + \sum_{l = j+b}^{i-1} p[l] \\
 \sum_{l=j}^{i-\tau} k_l &= \sum_{l = j+b}^{i} p[l].
\end{align*}

\end{IEEEproof}

Next, we present Theorem~\ref{thm:osc_opt}, which shows that the $(\tau,b)$-\vgms is rate-optimal for \regimeNoSpread.

The proof involves an inductive argument on the time slot.
It will show that the cumulative number of symbols sent by each time slot under any code construction, even an offline one,
must be at least as many as under the $(\tau,b)$-\vgms to satisfy the \losslessDelay and \worstCaseDelay constraints.
{The proof technique synergizes with the greedy paradigm of the $(\tau,b)$-\vgms sending for each \messagePacket $\s[i]$:
(a) the minimal number of parity symbols needed to recover $\s[i]$ given any burst assuming that no future \messagePackets needs to be recovered, and (b) deferring the transmission of the parity symbols until the decoding deadline for $\s[i]$ (i.e., $\x[i+\delay]$).}
{The methodology for designing a streaming code using a greedy paradigm and inductively proving that it is rate-optimal form a suitable template for designing new online coding schemes in other regimes, as discussed in Section~\ref{sec:conclusion}.}
\begin{theorem}
\label{thm:osc_opt}
For any parameters $(\tau,b,\tau_L = 0)$, the ($\tau,b$)-\vgms is rate-optimal for transmission over a $C(b,\tau)$ channel.
\end{theorem}

\begin{IEEEproof}[Proof sketch]We present the full proof in
Appendix~\ref{sec:prove_thm_osc_opt}.

For an arbitrary \messageSizeSequence $k_0,k_1,\ldots,k_\len$, consider any optimal offline construction $O$. We prove by induction on time slot $i=0,1,2,\ldots,\len$ that the cumulative number of symbols sent by $O$ is at least as many as that of the $(\tau,b)$-\vgms.

In the base case, for each $i \in [\tau-1]$, the channel packet $X[i]$ under $O$ must contain at least $k_i$ symbols to meet the \losslessDelay constraint for \messagePacket $S[i]$.
Under the $(\tau,b)$-\vgms, $x[i]=k_i$.

The inductive step for $i \in \{\tau,\ldots,\len\}$ has two cases.

First, when no parity symbols are sent in $\x[i]$ (that is, $X[i]=S[i]$)
under the $(\tau,b)$-\vgms, at least $s[i]=k_i$ symbols are sent in $X[i]$ under $O$ to meet the \losslessDelay constraint.

Second, suppose that $X[i] = (S[i],P[i])$ is sent under the $(\tau,b)$-\vgms where $p[i]>0$.
Applying Lemma~\ref{lem:exactRec} shows that there is a burst loss starting at time slot $j \in \{i-\tau -b+1,\ldots,i-\tau\}$ where the number of parity symbols received under the $(\tau,b)$-\vgms in $X[j+b],\ldots,X[i]$ is the smallest for which it is possible to decode \messagePacket $S[j],\ldots,S[i-\tau]$.
We combine this fact with the \losslessDelay constraint for $S[j+b],\ldots,S[i]$.
We then show that at least as many symbols are sent under $O$ between time slots $(j+b)$ and $i$ as are, respectively, sent under the $(\tau,b)$-\vgms.
Applying the inductive hypothesis for time slot $(j+b-1)$ concludes the proof.

\end{IEEEproof}

We note that for any values of $\delay$ and $\burst$, the ($\tau,b$)-\vgms's rate (i.e., the optimal rate) is highly dependent on the precise sequence of the sizes of the messages. Hence, a closed-form expression is not viable.


Finally, we discuss \regimeTrivial (i.e., $\tau_L = (\tau-b)$ and $b|\tau$ ).
Under \regimeTrivial, {for any parameters $(\tau,b)$}, we show that a simple online coding scheme applied to each \messagePacket {has rate $\frac{\delay}{\delay+\burst}$.\footnote{The construction applies when $(\tau/b) | k_i$ for any $i \in [\len]$. This condition can be satisfied by padding each \messagePacket with up to $(\tau/b-1)$ symbols. For real-world live-streaming applications, the amount of padding is typically negligible (e.g., three orders of magnitude smaller than the average size of a \messagePacket).} Recall that $\frac{\delay}{\delay+\burst}$ is an upper bound on rate for the streaming model with variable-size \messagePackets~\cite{rudow2018variable}. Hence, the simple construction is rate-optimal.}

Under this encoding scheme, each \messagePacket $S[i]$ is evenly partitioned into $\frac{\tau}{b}$ components that are transmitted in channel packets $X[i],X[i+b],\ldots,X[i+\tau-b]$, respectively.
The parity symbols, in the form of the sum of these $\frac{\tau}{\burst}$ channel packets, are sent as $\x[i+\tau] = \sum_{j=0}^{\frac{\tau-\burst}{\burst}} \x[i+j\burst]$.\footnote{A generalized version of this construction appeared in~\cite{rudow2018variable} after the conference version~\cite{rudow2020Online} of our work included the construction presented here. A recent work employed a similar interleaving approach in designing a low complexity streaming code with linear field size in the setting of \textit{fixed-size message packets}~\cite{krishnan2019simple}. }
Note that in this coding scheme, each transmission occurs \textit{exactly} $\burst$ \channelPackets apart, which is only possible under \regimeTrivial. As such, each burst over $\x[i],\ldots,\x[i+\delay]$ drops precisely one of $X[i],X[i+b],\ldots,X[i+\tau-b],$ and $X[i+\tau]$. The remaining \channelPackets suffice to recover the missing one to meet the \worstCaseDelay constraint. Finally, we note that sending $\s[i]$ over $\x[i],\ldots,\x[i+\delay-\burst]$ satisfies the \losslessDelay constraint, as $(i+\delay_L) = (i+\delay-\burst)$.

{In this section, we presented rate-optimal online streaming codes for \regimeNoSpread and \regimeTrivial.
We showed in the proof of Theorem~\ref{thm:osc_opt} that, for any $(\tau,\burst)$, the $(\tau,\burst)-$\vgms matches the rate of the best offline construction possible for \regimeNoSpread.
The simple construction for \regimeTrivial matches the upper bound of the rate of $\frac{\delay}{\delay+\burst}$.
Both of these constructions match the best possible rates of the offline setting, establishing that the \onop equals the \offop in both parameter regimes.
The construction for \regimeNoSpread can be used for \textit{any} value of $\delay_L$, although it is not necessarily rate-optimal for $\delay_L>0$.
Next, in Section~\ref{sec:gap}, we show that online codes cannot match the \offop for all other parameter settings.}

\section{Infeasiblity of \offop for Online Schemes}

\label{sec:gap}

In Section~\ref{sec:achieve}, we presented online code constructions that matched the \offop under the two broad settings of  \regimeNoSpread and \regimeTrivial. A natural question is whether there are any other parameter settings where an online coding scheme can attain the \offop.
In this section, we show that the \onop is strictly less than the \offop for all other parameter settings.

At a high level, the optimal approach to spreading symbols from a \messagePacket $S[i]$ over channel packets $X[i],\ldots,X[i+\tau_L]$ depends on the sizes of future \messagePackets (i.e., $k_{i+1},\ldots,k_\len$).
This dependency enables offline coding schemes to have higher rates than online coding schemes in all settings besides \regimeNoSpread and \regimeTrivial,  as we will show in Theorem~\ref{thm:gap}.

\begin{theorem}
\label{thm:gap}
For any parameters $(\tau,b,\tau_L)$ outside of \regimeNoSpread and \regimeTrivial, the \onop is strictly less than \offop.

\end{theorem}

\begin{IEEEproof}[Proof sketch]
The proof consists of three mutually exclusive cases shown {via illustrative examples in Sections~\ref{sec:conv1Ex},~\ref{sec:conv2Ex}, and~\ref{sec:conv3Ex} and in detail in Appendix~\ref{sec:conv1},~\ref{sec:conv2}, and~\ref{sec:conv3}}. In each case, we present two distinct \messageSizeSequences of length $(\len+1)$, which match for the first several time slots. We show a lower bound on the \offop for the two \messageSizeSequences by presenting an offline coding scheme with rates $R_\len^{(1)}$ and $R_\len^{(2)}$ on the first and second \messageSizeSequences, respectively. To attain a rate of at least $R_\len^{(1)}$ on the first \messageSizeSequence requires sending symbols in a manner that leads to a lower rate than $R_\len^{(2)}$ on the second.
\end{IEEEproof}
\begin{remark}
Although Theorem~\ref{thm:gap} is proven for two specific \messageSizeSequences, a similar proof holds if the sizes of the \messagePackets were only approximately the sizes corresponding to the \messageSizeSequences. As such, the result establishes a broad class of \messageSizeSequences for which there is a gap between the \onop and the \offop.
\end{remark}

\subsection{Case $\tau_L \ge b $ and $\tau_L = (\tau-b)$ }
\label{sec:conv1Ex}
This section presents the proof for parameters $(\burst, \tau_L,\tau) = (3,4,7)$; the general case, which builds closely on this example, is proven in Appendix~\ref{sec:conv1}.

\begin{figure}
    \centering
    \includegraphics[height=50pt]{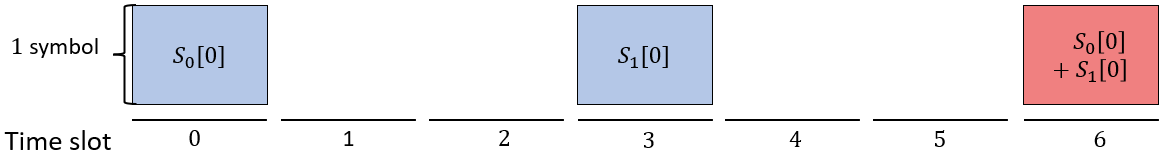}
    \caption{Offline construction for \messageSizeSequence $1$ for parameters $(b,\tau_L,\tau) = (3,4,7)$.}
    \label{fig:conv1Scheme1Toy}
\end{figure}

\begin{figure}
    \centering
    \includegraphics[height=80pt]{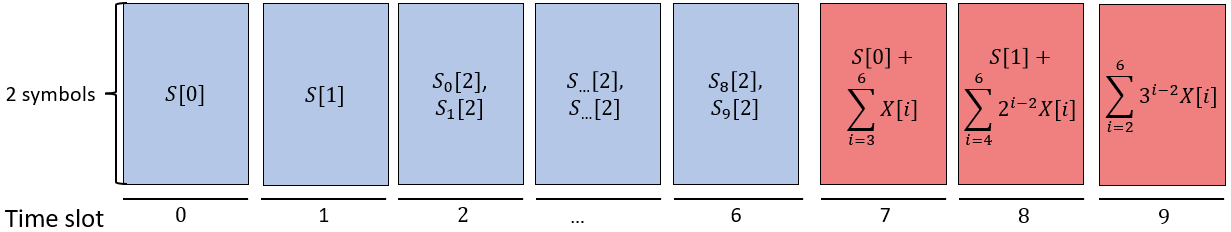}
    \caption{Offline construction for \messageSizeSequence $2$ for parameters $(b,\tau_L,\tau) = (3,4,7)$. }
    \label{fig:conv1Scheme2Toy}
\end{figure}

Consider the following two \messageSizeSequences:
\begin{enumerate}
    \item $k^{(1)}_0 =2$ and $k^{(1)}_j = 0$ for $j>0$.
    \item $k^{(2)}_0 =2, k^{(2)}_1 =2,k^{(2)}_2 =10,$ and $k^{(1)}_j = 0$ for $j>2$.
\end{enumerate}
An offline construction for the two \messageSizeSequences is shown in Figures~\ref{fig:conv1Scheme1Toy} and~\ref{fig:conv1Scheme2Toy} respectively, over $\F_{q}$ for any prime $q \ge 83$.

For \messageSizeSequence $1$, the construction sends $\x[0] = \s_0[0]$, $\x[3] = \s_1[0]$, and $\x[6] = \left( \s_0[0] + \s_1[0]\right)$, as shown in Figure~\ref{fig:conv1Scheme1Toy}. The \losslessDelay constraint is trivially satisfied. The \worstCaseDelay constraint is met, as at most one of $\x[0],\x[3],$ and $\x[6]$ is lost.

For \messageSizeSequence $2$, the construction sends $\x[0] = \s[0], \x[1] = \s[1]$, for $i \in \{2,\ldots,6\}$ sends $\x[i] = \left(\s_{2(i-2)}[2], \s_{2(i-2)+1}[2]\right)$, $\x[7] = \left(\s[0] + \sum_{i=3}^6 \x[i]\right), \x[8] = \left(\s[1] + \sum_{i=4}^6 2^{i-2} \x[i]\right)$, and $\x[9] = \sum_{i=2}^6 3^{i-2} \x[i]$, as shown in Figure~\ref{fig:conv1Scheme2Toy}. The \losslessDelay is clearly satisfied. The \worstCaseDelay constraint is met, as will be shown next through a comprehensive case analysis. For any $l \in \{0,1\}$ suppose that $\x[l]$ is lost, then $\s[l] = \left(\x[7+l] -\sum_{j=3+l}^6 (l+1)^{j-2}\x[j]\right)$ is obtained within $7$ time slots. When $\x[2]$ is lost, $\s[0]$ and $\s[1]$ are decoded. Then one can decode
\begin{align*}
 \left(\s_4[2], \s_5[2]\right) &= 2^{-2}\left(\x[8] - \s[1] -2^3\x[5]-2^4\x[6]\right) \\
\left(\s_2[2], \s_3[2]\right) &= \left( \x[7] - \s[0] - \x[4] - \x[5] - \x[6]\right) \\
\left(\s_0[2], \s_1[2]\right) &= \left(\x[9] - \sum_{j=3}^6 3^{j-2}\x[j] \right).
\end{align*}
When a burst starts with $\x[3]$, $\s[0],\s[1],$ and $\s_0[2]$ are decoded, and $\left(\s_8[2], \s_9[2]\right)$ is received.
Combining $\s[0],\s[1],$ and $\x[2],$ with $\x[6:9]$ yields $\sum_{l=3}^6 \x[i], \sum_{l=4}^5 2^{l-2} \x[l],$ and $\sum_{l=3}^5 3^{l-2} \x[l]$. These three equations are linearly independent and yield $\x[3:5]$.
Thus, $\s[2]$ is decoded by time slot $9$.
When a burst starts with $\x[4]$, $\s[0],\s[1], \s_0[2],\s_1[2],\s_2[2],$ and $\s_{3}[2]$ are received and combined with $\x[7],\x[8],$ and $\x[9]$ to determine $\sum_{j = 4}^6 \x[j], \sum_{j = 4}^6 2^{j-2} \x[j]$, and $\sum_{j = 4}^6 3^{j-2} \x[j]$.
These three equations are linearly independent and yield $\x[4:6]$, which consist of $\s_4[2],\ldots,\s_9[2]$.
When a burst starts with $\x[5]$, $\s[1], \s_0[2],\ldots,\s_5[2]$ are received and combined with $\x[8]$ and $\x[9]$ to determine $\sum_{j = 5}^6 2^{j-2} \x[j]$, and $\sum_{j = 5}^6 3^{j-2} \x[j]$.
These two equations are linearly independent and yield $\x[5]$ and $\x[6]$, which include $\s_6[2],\ldots,\s_9[2]$.
When a burst starts with $\x[6]$, $\s_0[2],\ldots,\s_{7}[2]$ are received, leading to $\left(\s_8[2],\s_9[2]\right) = 3^{-4}\left( \sum_{j=2}^5 3^{j-2} \x[j]\right)$.
When $\x[0:6]$ are received, the \messagePackets are received.

The rate of the offline construction for \messageSizeSequence $1$ is $2/3$, while its rate for \messageSizeSequence $2$ is $0.7$.
An online construction must send at most $1$ symbol in $\x[0]$ to have a rate of $2/3$ on \messageSizeSequence $1$ because $\x[0]$ can be lost.
We next show that any such scheme cannot attain the rate of $0.7$ on \messageSizeSequence $2$.
If \messageSizeSequence $2$ occurs, the online construction must send at least $13$ symbols over $\x[1:6]$ due to the \losslessDelay constraint.
At least one of $\x[1:3]$ and $\x[4:6]$ must contain at least $7$ symbols and may be lost.
At least $14$ symbols must be received. So the rate is at most $14/21$ (i.e., less than $0.7$).
Therefore, any online construction with a rate of $2/3$ on \messageSizeSequence $1$ cannot attain the rate of $0.7$ on \messageSizeSequence $2$, unlike the proposed offline construction.

\subsection{Case $\tau_L < b$ and $\tau_L= (\tau-b)$}
\label{sec:conv2Ex}
This section presents the proof for parameters $(\burst, \tau_L,\tau) = (2,1,3)$; the general case, which builds closely on this example, is proven in Appendix~\ref{sec:conv2}.

\begin{figure}
    \centering
    \includegraphics[height=80pt]{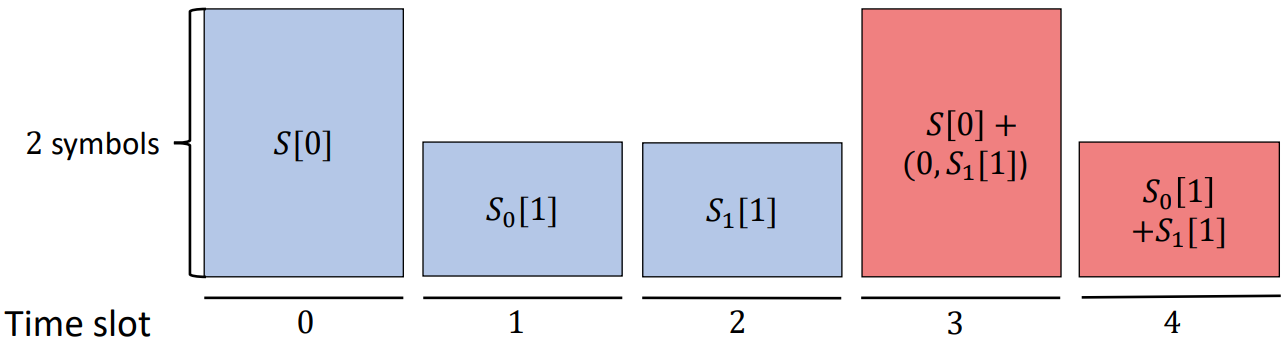}
    \caption{Offline construction for \messageSizeSequence $1$ for parameters $(b,\tau_L,\tau) = (2,1,3)$.}
    \label{fig:conv2Scheme1Toy}
\end{figure}

\begin{figure}
    \centering
    \includegraphics[height=80pt]{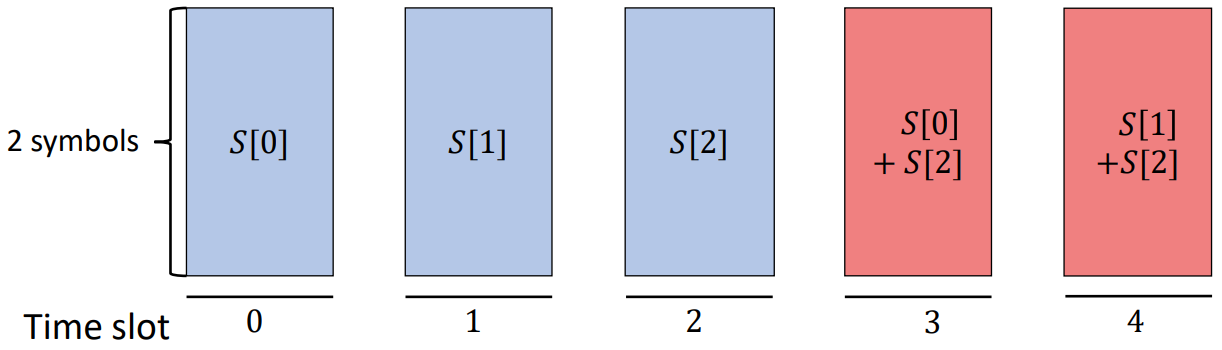}
    \caption{Offline construction for \messageSizeSequence $2$ for parameters $(b,\tau_L,\tau) = (2,1,3)$.}
    \label{fig:conv2Scheme2Toy}
\end{figure}

{Consider the following two \messageSizeSequences}:
\begin{enumerate}
    \item $k^{(1)}_0 =2,$ $k^{(1)}_1 =2$, and $k^{(1)}_j = 0$ for $j>1$.
    \item $k^{(1)}_0 =2,$ $k^{(1)}_1 =2$, $k^{(1)}_2 =2$, and $k^{(1)}_j = 0$ for $j>2$.
\end{enumerate}
{An offline construction for the two \messageSizeSequences is shown in Figures~\ref{fig:conv2Scheme1Toy} and~\ref{fig:conv2Scheme2Toy} respectively over any finite field, $\F_q$}.

For \messageSizeSequence $1$, the construction sends $\x[0] = \s[0]$, $\x[1] = \s_0[1]$, $\x[2] = \s_1[1]$, $\x[3] = \left(\s[0] + (0, \s_1[1]) \right),$ and $\x[4] = \left( \s_0[1] + \s_1[1]\right)$, as shown in Figure~\ref{fig:conv2Scheme1Toy}. The \losslessDelay constraint is trivially satisfied. The \worstCaseDelay constraint is met for $\s[0]$ because either $\s[0]$ is received, or $\s_1[1]$ and $\x[3]$ are received, yielding $\s[0]$.
When $\x[1]$ is lost, $(0,\s_1[1]) = \left(\x[3] - \s[0]\right)$ is obtained, leading to $\s_0[1] = \left( \x[4] - \s_1[1]\right)$.
When $\x[2]$ is lost, $\s_0[1]$ is decoded, leading to $\s_1[1] = \left( \x[4] - \s_0[1]\right)$.
As such, the \worstCaseDelay is satisfied for $\s[1]$.

For \messageSizeSequence $2$, the construction sends $\x[0] = \s[0], \x[1] = \s[1]$, $\x[2] = \s[2]$, $\x[3] = \left( \s[0] + \s[2]\right),$ and $\x[4] = \left(\s[1] + \s[2]\right)$, as shown in Figure~\ref{fig:conv2Scheme2Toy}. The \losslessDelay is clearly satisfied. The \worstCaseDelay constraint is met for $\s[0]$ as either $\x[0]=\s[0]$ is received, or $\s[0]=\left(\x[3] - \x[2]\right)$ is obtained.
The \worstCaseDelay constraint is satisfied for $\s[1]$ since either $\x[1]$ is received, or $\s[0]$ is decoded, leading to $\s[2] = \left(\x[3] - \s[0]\right)$, and $\s[1] = \left(\x[4] - \s[2]\right)$.
The \worstCaseDelay constraint is satisfied for $\s[2]$ because either $\x[2]$ is received, or $\s[1]$ is decoded, yielding $\s[2] = \left(\x[4] - \s[1]\right)$.

The offline construction's rate for \messageSizeSequence $1$ is $4/7$, while its rate for \messageSizeSequence $2$ is $0.6$.
An online construction with a rate of $4/7$ on \messageSizeSequence $1$ must send at most $3$ symbols in $\x[0:1]$, since at least $4$ symbols are sent in $\x[2:4]$ in case $\x[0:1]$ is lost.
Also, the construction sends at least $2$ symbols over $\x[0:1]$ to recover $\s[0]$ under lossless transmission.
Next, we show that any such scheme cannot attain the rate of $0.6$ on \messageSizeSequence $2$ due to sending fewer than $4$ symbols over $\x[0:2]$.
Thus, any online construction with a rate of $4/7$ on \messageSizeSequence $1$ cannot attain the rate of $0.6$ on \messageSizeSequence $2$, unlike the proposed offline construction.

First, suppose that exactly $2$ symbols are sent in $\x[0:1]$.
Then $\x[2:3]$ suffices to recover $\s[0]$.
Recall that the $2$ symbols in $\x[0:1]$ only contain information about $\s[0]$, as they suffice to recover $\s[0]$ under a lossless transmission.
Thus, $\x[0:1]$ are recovered as a function of $\s[0]$, leaving the transmission lossless, so $\s[1:2]$ are recovered.
Thus, $\x[2:3]$ contains at least $6$ symbols.
At least $6$ symbols are sent outside of $\x[2:3]$ in case $\x[2:3]$ is lost, so the rate is at most $6/12$.

Second, due to the upper bound on the rate of $\frac{\delay}{\delay+\burst} = \frac{3}{5}$ and \worstCaseDelay, at least $10 = 6*\frac{5}{3}$ symbols must be sent by time slot $5$. Suppose exactly $3$ symbols are sent in $\x[0:1]$.
Consider the $5$ periodic erasure channels, $C_0,\ldots,C_4$, where for $i \in [4]$, $C_i$ drops packets $\x[j:j+1]$ for all $j \equiv i \mod 5$.
Each packet is dropped by $2$ of these channels, so the channels drop at least $\frac{2}{5}*10 \ge 4$ symbols on average.
At least $6$ symbols must be received to ensure recovery.
If any channel dropped $5$ or more symbols, the rate would be at most $6/11$.
Thus, each channel must drop exactly $4$ symbols to attain a rate of $0.6$.
Therefore, $C_0$ drops exactly $4$ symbols\textemdash $3$ over $\x[0:1]$ and $1$ in $\x[5]$.
Each of $C_4, C_3,$ and $C_2$ must drop $4$ symbols (i.e., $n_4 + n_5 = 4, n_3+n_4 = 4, n_2+n_3 = 4$).
Hence, $\x[4]$ contains $3$ symbols, $\x[3]$ contains $1$ symbol, and $\x[2]$ contains $3$ symbols.
In total, $(3 + 3 + 1 + 3+1) = 11$ symbols are sent over $\x[0:1], \x[2], \x[3],\x[4],$ and $\x[5]$, leading to a rate of $6/11$, which is less than $0.6$.

Therefore, any online construction that matches the rate of $4/7$ on \messageSizeSequence $1$ cannot attain the rate of $0.6$ on \messageSizeSequence $2$, unlike the offline construction.

\subsection{Case $\tau_L < (\tau-b)$}
\label{sec:conv3Ex}
This section presents the proof for parameter $(\burst, \tau_L,\tau) = (1,1,3)$; the general case, which builds closely on this example, is proven in Appendix~\ref{sec:conv3}.

\begin{figure}
    \centering
    \includegraphics[height=50pt]{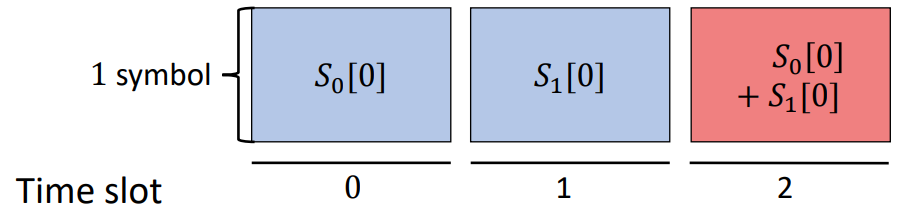}
    \caption{Offline construction for \messageSizeSequence $1$ for parameters $(b,\tau_L,\tau) = (1,1,3)$.}
    \label{fig:conv3Scheme1Toy}
\end{figure}

\begin{figure}
    \centering
    \includegraphics[height=85pt]{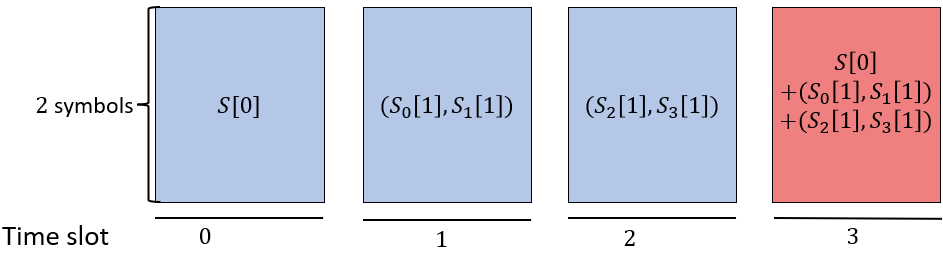}
    \caption{Offline construction for \messageSizeSequence $2$ for parameters $(b,\tau_L,\tau) = (1,1,3)$.}
    \label{fig:conv3Scheme2Toy}
\end{figure}

Consider the following two \messageSizeSequences:
\begin{enumerate}
    \item $k^{(1)}_0 =2$ and $k^{(1)}_j = 0$ for $j>0$.
    \item $k^{(1)}_0 =2,$ $k^{(1)}_1 =4$, and $k^{(1)}_j = 0$ for $j>1$.
\end{enumerate}
An offline construction for the two \messageSizeSequences is shown in Figures~\ref{fig:conv3Scheme1Toy} and~\ref{fig:conv3Scheme2Toy} respectively over any finite field, $\F_q$.

For \messageSizeSequence $1$, the construction sends $\x[0] = \s_0[0]$, $\x[1] = \s_1[0]$, and $\x[2] = \left(\s_0[0]+\s_1[0]\right)$, as is shown in Figure~\ref{fig:conv3Scheme1Toy}. The \losslessDelay constraint is trivially satisfied. The \worstCaseDelay constraint is met because at most one of $\x[0],\x[1],$ or $\x[2]$ is lost and $\x[2] = \left(\x[0] + \x[1]\right)$.

For \messageSizeSequence $2$, the construction sends $\x[0] = \s[0], \x[1] = \left(\s_0[1], \s_1[1] \right)$, $\x[2] = \left( \s_2[1], \s_3[1]\right)$, and $\x[3] = \left( \s[0] + (\s_0[1], \s_1[1]) + (\s_2[1], \s_3[1])\right)$, as shown in Figure~\ref{fig:conv3Scheme2Toy}. The \losslessDelay is clearly satisfied. The \worstCaseDelay constraint is met, since at most one of $\x[0],\x[1],\x[2],$ or $\x[3] = \sum_{i=0}^2 \x[i]$ is lost.

The offline construction's rate for \messageSizeSequence $1$ is $2/3$, while its rate for \messageSizeSequence $2$ is $0.75$.
For an online construction to attain a rate of $2/3$ on \messageSizeSequence $1$, it must send exactly $1$ symbol in each of $\x[0]$ and $\x[1]$ due to (a) the \losslessDelay constraint and (b) ensuring at most $1$ symbol is lost\textemdash a necessity to attain the rate of $2/3$.
Next, we show that any such scheme cannot attain the rate of $0.75$ on \messageSizeSequence $2$.
If \messageSizeSequence $2$ occurs, at least $6$ symbols are sent over $\x[0:2]$ due to the \losslessDelay constraint. The average number of symbols per packet is at least $2$.
If $\x[0]$ contains one symbol, at least one of $\x[1]$ or $\x[2]$ contains at least $3$ symbols.
At least $6$ symbols must be received to satisfy the \worstCaseDelay constraint. Since at least $3$ symbols may be lost, at least $9$ symbols must be sent in total.
As such, the rate is at most $2/3$, which is less than $0.75$.
Therefore, any online construction that matches the rate of $2/3$ on \messageSizeSequence $1$ cannot attain the rate of $0.75$ on \messageSizeSequence $2$, unlike the offline construction.

\section{Conclusion}
\label{sec:conclusion}
Real-time streaming applications, such as videoconferencing, transmit a sequence of messages of varying sizes.
These applications operate in an online setting without access to future message sizes.
However, previously studied upper bounds on the rate apply to an offline setting with advance access to the sizes of all messages, leaving the best possible rate of the online setting an open question.
We introduce the \textit{first rate-optimal online coding schemes} for two broad parameter regimes (that is, \regimeNoSpread and \regimeTrivial) which are optimal even for the offline setting.
To do so, we propose a framework for designing online constructions using a greedy paradigm for sending parity symbols and inductively analyzing the rate that is suitable for future works for \messagePackets of varying sizes.
We also show for all other parameter regimes that the best way to spread the symbols of messages over multiple transmissions depends on the sizes of future messages.
Consequently, \textit{no online coding scheme can match the optimal rate of offline coding schemes}.

The gap between the \onop and \offop prompts three directions of further study for the {parameter settings outside of \regimeNoSpread and \regimeTrivial}.
First, how can one design rate-optimal offline code constructions?
Second, what does it mean to be rate-optimal in the online setting, given that the rate depends on the specific sequence of sizes of future messages?
{Third, can one use the proposed methodology to design and analyze online constructions to design rate-optimal or approximately rate-optimal online streaming codes?}
These questions have been partially answered for the smallest \losslessDelay where spreading message symbols can alleviate the variability of the sizes of \messagePackets (i.e., $\delay_L=1$) in~\cite{rudow2022learning}; the questions remain open for large values of $\delay_L$.

\appendix

\label{sec:appendix}
\subsection{Proof of Theorem~\ref{thm:osc_opt}}
\label{sec:prove_thm_osc_opt}
In this section, we will prove Theorem~\ref{thm:osc_opt}. At a high level, the proof is inductive and shows that the cumulative number of symbols sent by each time slot under the $(\tau,b)$-\vgms is the minimum possible.
For time slots where no parity symbols are sent, it follows immediately by the \losslessDelay constraint.
Otherwise, there is some burst for which every parity symbol in the received \channelPackets is needed to recover the burst within the \worstCaseDelay.

We begin by introducing the preliminary notation for the proof. We then include a few auxiliary Lemmas used throughout the proof. Finally, we present the full proof itself.

Let $\len$ be an arbitrary natural number{, and consider any length $(\len+1)$ \messageSizeSequence $k_0,\ldots,k_\len$}. Let $O$ be an arbitrary offline code construction that satisfies the \losslessDelay and \worstCaseDelay constraints over a $C(b,\tau)$ channel for the \messageSizeSequence.
Let the channel packet transmitted during time slot $j\in [\len]$ under construction $O$ and under the $(\tau,b)$-\vgms be labeled as $X_O[j]$ and $X_V[j]$, respectively. Let the cumulative number of symbols transmitted through time slot $j$ under construction $O$ and under the $(\tau,b)$-\vgms be denoted $n^+_{O,j} = \sum_{i=0}^j x_O[i]$  and $n^+_{V,j} = \sum_{i=0}^j x_V[i]$, respectively.
Recall from Section~\ref{sec:modBack} that each \messagePacket comprises symbols drawn independently and uniformly at random from the finite field $\F_q$.
Let $\mathcal{S}$ be a random variable representing a uniformly random element of $\F_q$.

Next, we show that the \losslessDelay constraint necessitates transmitting at least as many symbols as the size of the \messagePacket for each time slot.
\begin{lemma}
\label{lem:sendK}
Consider any parameters $(\tau,b,\delay_L = 0)$, an arbitrary \messageSizeSequence $k_0,k_1,\ldots,k_\len$, and any code construction which satisfies the \losslessDelay and \worstCaseDelay constraints over a $C(b,\tau)$ channel. For any $j \in [\len]$, $n_j \ge k_j$.
\end{lemma}
\begin{IEEEproof}
Follows directly from (a) the independence of \messagePackets, and (b) the \losslessDelay constraint for $\tau_L = 0$.
\end{IEEEproof}

Next, we establish that whenever a burst of length $b$ occurs, all \messagePackets from time slots before the burst must be decoded before the burst to satisfy both the \losslessDelay and \worstCaseDelay constraints.

\begin{lemma}
\label{lem:decodeBeforeBurst}
Consider any parameters $(\tau,b,\delay_L = 0)$, an arbitrary \messageSizeSequence $k_0,k_1,\ldots,k_\len$, $j \in [\len]$, and any code construction which satisfies the \losslessDelay and \worstCaseDelay constraints over a $C(b,\tau)$ channel. When $X[j],\ldots,X[j+b-1]$ are lost in a burst, $\s[0:j-1]$ are decoded by time slot $(j-1)$.
\end{lemma}
\begin{IEEEproof} By the \worstCaseDelay constraint, $S[0:j-\tau-1]$ are all decoded by time slot $(j-1)$.
Under the $C(b,\tau)$ channel, when $\x[j],\ldots,\x[j+\burst-1]$ are lost, $X[j-\tau],\ldots,X[j-1]$ are necessarily received.\footnote{When $j < \tau,$ $\x[0:j-1]$ are received.}
By the \losslessDelay constraint, $\s[0:j-\delay-1]$ and $\x[j-\delay:j-1]$ suffice to decode $\s[j-\delay:j-1]$.
\end{IEEEproof}

Finally, we prove Theorem~\ref{thm:osc_opt} below.

\begin{IEEEproof}[Proof of Theorem~\ref{thm:osc_opt}]Let $k_0,k_1,\ldots,k_\len$ be an arbitrary \messageSizeSequence.
We will show by induction that the cumulative number of symbols sent through time slot $i\in [\len]$ under an arbitrary offline construction, $O$, is at least as many as that of the $(\tau,b)-$\vgms (i.e., $n^+_{O,i} \ge n^+_{V,i}$).
Consequently, the $(\tau,b)$-\vgms matches the \offop.

In the base case, we consider $j \in [\tau-1]$.
Applying Lemma~\ref{lem:sendK} determines that $x_O[j] \ge k_j = x_V[j]$ $\forall j \in [\tau-1]$.

For the inductive hypothesis, we assume that for some $(i_* \ge \tau-1)$, for all $l \in [i_*]$, $n^+_{O,l} \ge n^+_{V,l}$.

For the inductive step, consider the time slot $(i = i_*+1 \ge \tau)$. By the inductive hypothesis, $n^+_{O,i-1} \ge n^+_{V,i-1}$. We will show that $n^+_{O,i} \ge n^+_{V,i}$ using two cases.

\textbf{Case  $x_V[i] = k_i$}:

Applying Lemma~\ref{lem:sendK} determines that $x_O[i] \ge k_i$. Therefore, $(n^+_{O,i} =n^+_{O,i-1}+k_i \ge n^+_{V,i-1}+k_i = n^+_{V,i})$.

\textbf{Case $x_V[i] > k_i$}:
We first provide a high-level intuition of the proof and then the detailed derivation. \\
\textit{High-level summary:}  Applying Lemma~\ref{lem:exactRec} shows that there is a burst starting in time slot $j \in \{i-\tau -b+1,\ldots,i-\tau\}$ for which the $(\tau,b)$-\vgms receives minimum required number of parity symbols to decode \messagePackets $\s[j:i-\delay]$ by time slot $i$.
Combining this fact with meeting the \losslessDelay constraint for $\s[j+\burst:i]$ shows that the number of symbols sent under $O$ between time slots $(j+b)$ and $i$ is at least as many as that of the $(\tau,b)$-\vgms.

\textit{Detailed derivation:}
By Lemma \ref{lem:exactRec}, there is some $j \in \{i-\tau-b+1,\ldots,i-\tau\}$ such that $\sum_{l=j+b}^{i} p[l] = \sum_{l=j}^{i-\tau} k_l$. Therefore,
\begin{equation}
\label{eq:equality}
\sum_{l=j+b}^{i} x_V[l] = \sum_{l=j}^{i-\tau} k_l+\sum_{l=j+b}^i k_{l}.
\end{equation}

Next, we show that at least as many symbols are sent over $\x_O[j+\burst:i]$ as are sent over $\x_V[j+\burst:i].$
Consider a burst loss of $X[j],\ldots,X[j+b-1]$.
Applying Lemma~\ref{lem:decodeBeforeBurst} shows that $\s[0:j-1]$ are known by the receiver by time slot $(j-1)$.
By the \worstCaseDelay constraint,
\begin{equation}
\label{eq:recWorst}
H\left(\s[j:i-\delay]\big |\x_O[j+\burst:i],\s[0:j-1]\right)=0.
\end{equation}

We next bound the number of symbols sent over $\x_O[j+\burst:i]$ ass
\begin{align}
H\left(\s[j:i-\delay]\right) + H\left(\x_O[j+\burst:i]\big |\s[0:i-\delay]\right) =& H\left(\x_O[j+\burst:i], \s[j:i-\delay]\big |\s[0:j-1]\right) \label{eq:Qcond1}\\
=& H\left(\x_O[j+\burst:i]\big |\s[0:j-1]\right) \label{eq:Qcond2}\\
    +& H\left( \s[j:i-\delay]\big |\s[0:j-1], \x_O[j+\burst:i]\right) \nonumber\\
=&H\left(\x_O[j+\burst:i]\big |\s[0:j-1]\right)\label{eq:Qcond3},
\end{align}
{where Equation~\ref{eq:Qcond1} follows from the chain rule and independence of \messagePackets, Equation~\ref{eq:Qcond2} follows from the chain rule, and Equation~\ref{eq:Qcond3} follows from Equation~\ref{eq:recWorst}.}

Combining Equations~\ref{eq:Qcond1} and~\ref{eq:Qcond3} with the fact that conditioning reduces entropy yields
\begin{equation}
\label{eq:ineqOverall}
H\left(\x_O[j+\burst:i]\right) \ge H\left(\x_O[j+\burst:i]\big |\s[0:j-1]\right) \ge H\left(\s[j:i-\delay]\right) + H\left(\x_O[j+\burst:i]\big |\s[0:j+\burst-1]\right).
\end{equation}

{Next, we evaluate the size of $H\left(\x_O[j+\burst:i]\big |\s[0:j+\burst-1]\right)$ as
\begin{align}
H\left(\s[j+\burst:i],\x_O[j+\burst:i]\big |\s[0:j+\burst-1]\right) =& H\left(\s[j+\burst:i]\right) + H\left(\x_O[j+\burst:i]\big |\s[0:i]\right)\label{eq:afterBurst1}\\
=&H\left(\s[j+\burst:i]\right) \label{eq:afterBurst2}\\
=&H\left(\x_O[j+\burst:i]\big |\s[0:j+\burst-1]\right) + \label{eq:afterBurst3}\\
&H\left(\s[j+\burst:i]\big |\s[0:j+\burst-1],\x_O[j+\burst:i]\right) \nonumber\\
=&H\left(\x_O[j+\burst:i]\big |\s[0:j+\burst-1]\right) \label{eq:afterBurst4},
\end{align}
where Equation~\ref{eq:afterBurst1} follows from conditioning and independence of \messagePackets, Equation~\ref{eq:afterBurst2} follows from the fact that for $l \in [\len]$, $\x_O[l]$ is a function of $\s[0:l]$, Equation~\ref{eq:afterBurst3} follows from conditioning, and Equation~\ref{eq:afterBurst4} follows from the \losslessDelay constraint.}

{For any $i \in [\len]$,
\begin{equation}
\label{eq:sizeMessage}
H(\s[i]) = H(\mathcal{S})k_i
\end{equation}
\begin{equation}
\label{eq:sizeChannel}
H(\x[i]) \le H(\mathcal{S})n_i
\end{equation}
where $\mathcal{S}$ was defined as a random variable drawn uniformly at random from the underlying field, $\F_q$.
This follows from the definition of \messagePackets, and the fact that the maximum possible entropy of $n_i$ symbols is $n_i H(\mathcal{S})$.}
Applying Equation~\ref{eq:afterBurst4} and~\ref{eq:afterBurst2} to Equations~\ref{eq:ineqOverall},~\ref{eq:sizeMessage}, and~\ref{eq:sizeChannel} yields
\begin{equation}
\label{eq:ineqFinal}
        H(\mathcal{S}) \sum_{l = j+\burst}^{i} \big |\x_O[l]\big |\ge H\left(\x_O[j+\burst:i]\right) \ge H\left(\s[j:i-\delay]\right) +H\left(\s[j+\burst:i]\right) = H(\mathcal{S}) \big(\sum_{l = j}^{i-\delay} k_l + \sum_{l = j+\burst}^i k_l \big).
\end{equation}

Combining Equations~\ref{eq:ineqFinal} and~\ref{eq:equality} determines that
\begin{equation}
\label{eq:PassedBurst}
    H\left(\mathcal{S}\right) \sum_{l=j+b}^{i} x_O[l] \ge H\left(\mathcal{S}\right)\sum_{l=j+b}^{i} x_V[l].
\end{equation}

By definition, $(n^+_{O,i} = n^+_{O,j+b-1}+\sum_{l=j+b}^{i} x_O[l])$ and $(n^+_{V,i} = n^+_{V,j+b-1}+\sum_{l=j+b}^{i} x_V[l])$. Applying the inductive hypothesis to $(j+b-1<i)$ shows that $(n^+_{V,j+b-1} \le n^+_{O,j+b-1})$. Combining the above equations with Equation~\ref{eq:PassedBurst} determines that $n^+_{V,i}  \le n^+_{O,i}$.
The inductive hypothesis is proven, and the result follows immediately.

\end{IEEEproof}

\subsection{Proof of Theorem~\ref{thm:gap} case $\tau_L \ge b$ and $\tau_L =(\tau-b)$}
\label{sec:conv1}
Let $(a = \lfloor \frac{\tau_L}{b}\rfloor)$ and $(e \equiv \tau_L \mod b)$. Theorem~\ref{thm:gap} does not apply when $\tau = (\tau_L-\burst)$ and $\burst | \tau$, necessitating that $(e>0)$.
Let $d$ be an arbitrary multiple of $(a+1)$.

Consider the following two \messageSizeSequences for which the offline construction will be shown below in Figures~\ref{fig:conv1Scheme1} and~\ref{fig:conv1Scheme2} respectively:
 \begin{enumerate}
     \item $ k^{(1)}_0 = \ldots = k^{(1)}_{e-1} = d$, and $k^{(1)}_e = \ldots = k^{(1)}_\len = 0$.
     \item $k^{(2)}_0 = \ldots = k^{(2)}_{b-2} = d$, $k^{(2)}_{b-1} = d(\tau_L+1),$ and $k^{(2)}_b = \ldots = k^{(2)}_\len = 0$.
 \end{enumerate}
Before going into the details of the proof,
we note that the proof applies for any value of $d$.
When $d$ is sufficiently large, the proof could also be extended to \messageSizeSequences where the \messagePackets' sizes may only approximately equal the ones in the \messageSizeSequence. More generally, the proof also applies for any \messageSizeSequences for which there is a subsequence of (a) $\delay$ \messagePackets whose sizes are $\ll d$, then (b) one of the two above \messageSizeSequences, then (c) another $\delay$ \messagePackets whose sizes are $\ll d$.

\begin{figure}
            \centering
\includegraphics[angle=0, height=60pt]{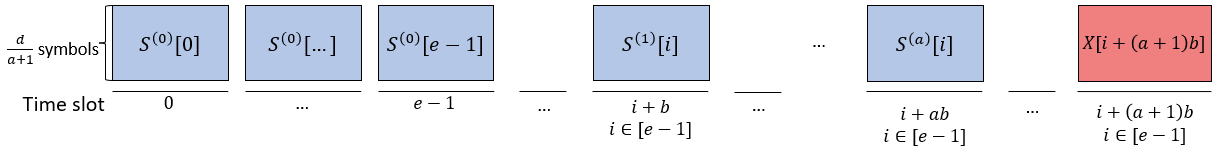}
\caption{The offline scheme for \messageSizeSequence $1$ for case $\tau_L \le b$ and $\tau_L =(\tau-b)$. Blue \channelPackets consist of message symbols, and red \channelPackets consist of parity symbols. The numbers under the lines at the bottom indicate the time slots. The offline scheme sends $\frac{d}{a+1}$ symbols in each of the first $e$ \channelPackets. }
\label{fig:conv1Scheme1}
\end{figure}
\begin{figure}
            \centering
        \includegraphics[angle=0, height=80pt]{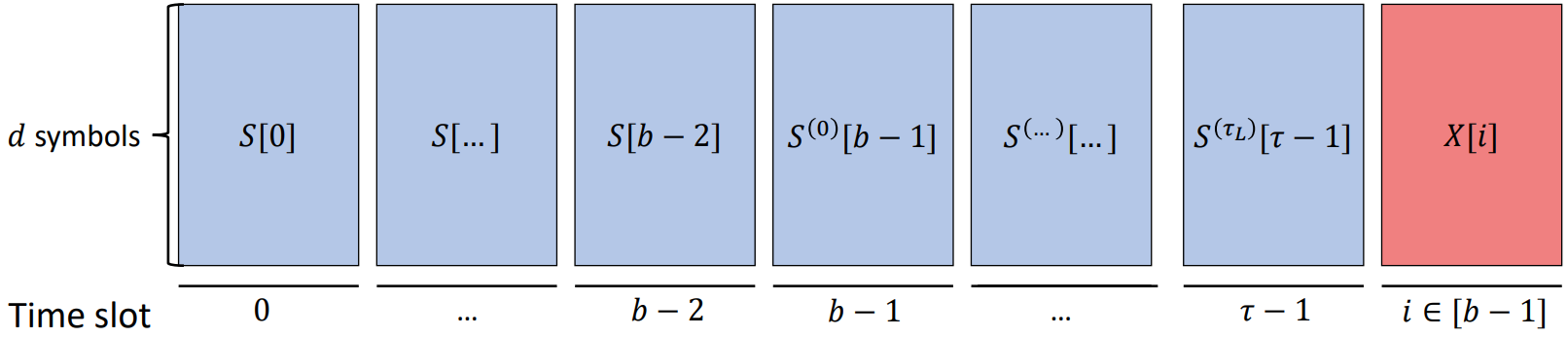}
        \caption{The offline scheme for \messageSizeSequence $2$ for case $\tau_L \le b$ and $\tau_L =(\tau-b)$.  Blue \channelPackets consist of message symbols and red \channelPackets consist of parity symbols. The numbers under the lines at the bottom indicate the time slots. The offline scheme sends $d$ symbols in each of the first $e$ \channelPackets. }
        \label{fig:conv1Scheme2}

\end{figure}

{We present an offline coding scheme for \messageSizeSequences $1$ and $2$, which has rates
\begin{equation}
\label{eq:ratesConv1}
R_\len^{(1)} = \frac{a+1}{a+2}, \quad \quad R_\len^{(2)} = \frac{\tau}{\tau+b}
\end{equation}
on the two \messageSizeSequences, respectively.
We describe and then validate the scheme for each \messageSizeSequence.}

\para{\textit{Offline scheme for \messageSizeSequence $1$:}}
Each \messagePacket is encoded separately with parameters $(\tau'=\lfloor \frac{\tau}{b}\rfloor b,b'=b,\tau_L'=\tau'-b)$ as described in Section~\ref{sec:achieve}, shown in Figure~\ref{fig:conv1Scheme1},
and detailed below.
\begin{itemize}
    \item For $i \in [e-1]$, $S[i]$ is evenly divided into $(a+1)$ components of size $d$ each: $S^{(0)}[i],\ldots,S^{(a)}[i]$. For $j \in [a], \x[i+j\burst] = \s^{(j)}[i]$.
    \item For $i \in [e-1]$, $X[i+(a+1)b] = \sum_{z =0}^aX[i+zb]$.
\end{itemize}

\noindent {\textbf{Decoding:}}
For $i \in [e-1]$, $S[i]$ is sent evenly over $\x[i],\x[i+b],\ldots,\x[i+ab]$ where $(i+ab) \le (i+\delay_L)$ and at most one of $X[i],X[i+b],\ldots,X[i+ab],$ or $X[i+(a+1)b] = \sum_{j=0}^a \x[i+jb]$ is lost.
Each \messagePacket is decoded within delay $\delay_L$ when the transmission is lossless and using a linear combination of the relevant $(a+1)$ \channelPackets within delay $\delay$ otherwise.

\para{\textit{Offline scheme for \messageSizeSequence $2$:}}
The first $(b-1)$ \messagePackets are sent with no delay and the symbols of the next \messagePacket are transmitted evenly over $X[b-1],\ldots,X[\tau-1]$.
The symbols of $\x[0:\tau-1]$ are used to create $d$ blocks of the rate $\frac{\tau}{\tau+\burst}$ systematic block code from~\cite{krishnan2019low}. Each of the $d$ blocks includes $\burst$ parity symbols that are sent in $\x[\tau],\ldots,\x[\tau+\burst-1]$ respectively.
The block code maps $\tau$ input symbols $(s_0,\ldots,s_{\tau-1})$ to $(\tau+b)$ codeword symbols $(s_0,\ldots,s_{\tau-1},p_0,\ldots,p_{b-1}).$
For each $j \in [\tau-1]$ and any burst erasing up to $b$ codeword symbols, the non-erased symbols of $(s_0,\ldots,s_{\tau-1},p_0,\ldots,p_{min(b-1,j)})$ are sufficient to decode $s_j$. Therefore, each symbol is recovered within $\tau$ symbols.
We note that although we use the block code from~\cite{krishnan2019low}, any other block code from~\cite{fong2018optimalLong,krishnan2018rate,dudzicz2019explicit,domanovitz2019explicit} also works.
The scheme is described in detail below {and shown in Figure~\ref{fig:conv1Scheme2}}:
\begin{itemize}
    \item  For $j \in [b-2]$, $X[j] = S[j]$.
    \item $S[b-1]$ is divided evenly into $(\tau_L+1)$ components of size $d$: $S^{(0)}[b-1],\ldots,S^{(\tau_L)}[b-1]$.
    \item For $j \in \{b-1,\ldots,\burst-1+\tau_L\},$ $\x[j] = S^{(j-\burst+1)}[\burst-1]$.
    \item For each $z \in [d-1]$, an instance of the block code from \cite{krishnan2019low} is created which maps $(X_z[0],\ldots,X_z[\tau-1])$ to $(X_z[0],\ldots,X_z[\tau-1],p_0^{(z)},\ldots,p_{b-1}^{(z)})$.
    \item For $j \in [b-1]$, $X[\tau+j] = (p_j^{(0)},\ldots,p_j^{(d-1)})$.
\end{itemize}

\noindent {\textbf{Decoding:}}
Each \messagePacket is transmitted within the current and next $\tau_L$ \channelPackets  and is, thus, decoded when the transmission is lossless. Each symbol $X_z[i]$ for $z \in [d-1]$ and $i \in [\tau-1]$ is decoded within the delay $\tau$ or by time slot $(\tau+b-1)$ using the block code $(X_z[0],\ldots,X_z[\tau-1],p_0^{(z)},\ldots,p_{b-1}^{(z)})$.
Hence, the \worstCaseDelay constraint is met.

\para{Proof of the converse result}: The \offop is at least $R_\len^{(1)}$ and $R_\len^{(2)}$ (that is, the rate of the offline scheme from Equation~\ref{eq:ratesConv1}) for \messageSizeSequences $1$ and $2$, respectively. Next, we show mutually exclusive conditions for the sum of the sizes of $X[0],\ldots,X[e-1]$ to have rates at least $R_\len^{(1)}$ and $R_\len^{(2)}$ on \messageSizeSequences $1$ and $2$ respectively. All online coding schemes, thus, fail the condition for at least one \messageSizeSequence since they are identical until time slot $e$.

\para{Condition for rate $R_\len^{(1)}$ on \messageSizeSequence $1$}: Consider any coding scheme for \messageSizeSequence $1$. At least $de$ symbols are sent over $X[b],\ldots,X[\len]$ since $X[0],\ldots,X[b-1]$ could be lost. At most $d\frac{e}{a+1}$ symbols can be sent over $X[0],\ldots,X[b-1]$ if the rate is at least $R_\len^{(1)}$.

\para{Condition for rate $R_\len^{(2)}$ on \messageSizeSequence $2$}: Consider an arbitrary coding scheme for \messageSizeSequence $2$.
At least $d\tau$ symbols are sent in $X[0],\ldots,X[\tau-1]$ to meet the \losslessDelay constraint. For each $i \in [a]$, at least $d\tau$ symbols are sent outside of $\x[e+ib:e+(i+1)b-1]$ in case $\x[e+ib:e+(i+1)b-1]$ is lost.
Since the rate is $R_\len^{(2)}$, at most $db$ symbols are sent in $\x[e+ib:e+(i+1)b-1]$.
As such, at least $(d\tau-d(a+1)b=de)$ symbols are sent in $X[0:e-1]$.

\para{Summary}:
Any online scheme whose rate is at least $R_\len^{(1)}$ on \messageSizeSequence $1$ sends at most $d\frac{e}{a+1}$ symbols in $X[0 : b-1]$.
As such, its rate is lower than $R_\len^{(2)}$ on \messageSizeSequence $2$.

\subsection{Proof of Theorem~\ref{thm:gap} case $\tau_L < b$ and $\tau_L =(\tau-b)$}

\label{sec:conv2}
Let $d$ be an arbitrary positive even integer.
Consider the following two \messageSizeSequences for which the offline construction will be shown below in Figures~\ref{fig:conv2Scheme1} and~\ref{fig:conv2Scheme2} respectively:
\begin{enumerate}
    \item $k^{(1)}_0 = \ldots = k^{(1)}_{b-\tau_L} = d$, and $k^{(1)}_{b-\tau_L+1} = \ldots = k^{(1)}_\len = 0$.
    \item $k^{(2)}_0 = \ldots = k^{(2)}_{b-\tau_L} = d$, $k^{(2)}_{b-\tau_L+1} = \ldots = k^{(2)}_{b-1} = 0$, $k^{(2)}_b = d$, and $k^{(2)}_{b+1} = \ldots = k^{(2)}_\len = 0$.
\end{enumerate}
{Before presenting the proof in detail, we observe that the proof could also be extended to similar \messageSizeSequences where the sizes of each \messagePacket is perturbed by a small amount as long as $d$ is large. More generally, the proof also applies to any \messageSizeSequence that contains one of the two above \messageSizeSequences proceeded and followed by $\tau$ \messagePackets sufficiently small relative to $d$.}

\begin{figure}
            \centering
\includegraphics[angle=0, height=85pt]{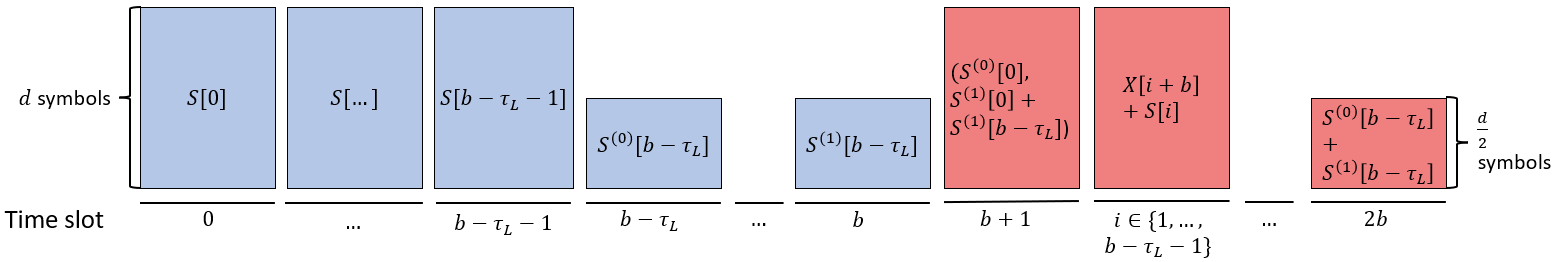}
\caption{{The offline scheme for \messageSizeSequence $1$ for case $\tau_L < \burst$ and $\tau_L =(\tau-b)$. Blue \channelPackets consist of message symbols, and red \channelPackets consist of parity symbols. The numbers under the lines at the bottom indicate the time slots. The offline scheme sends $\frac{d}{2}$ symbols in $\x[b-\delay_L]$.}}
\label{fig:conv2Scheme1}
\end{figure}
\begin{figure}
            \centering
        \includegraphics[angle=0, height=85pt]{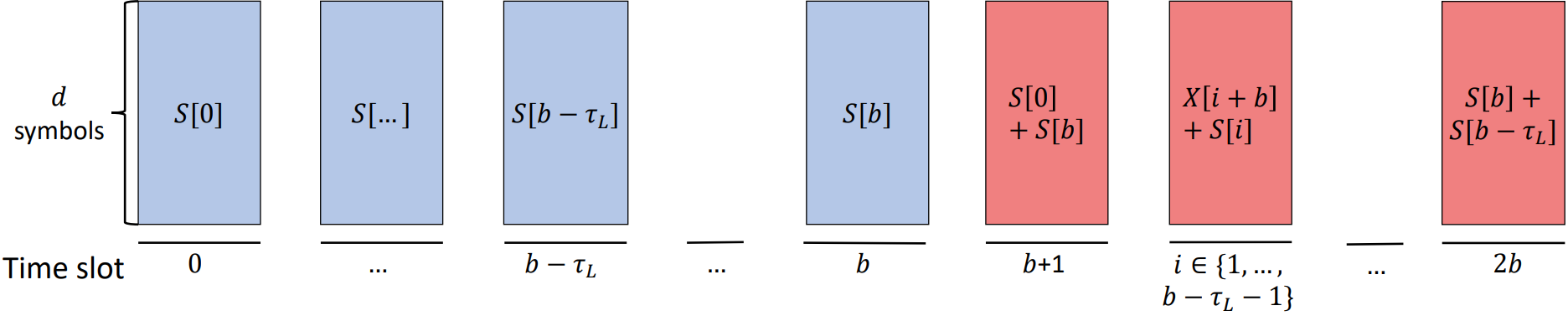}
        \caption{{The offline scheme for \messageSizeSequence $2$ for case $\tau_L < \burst$ and $\tau_L =(\tau-b)$. Blue \channelPackets consist of message symbols and red \channelPackets consist of parity symbols. The numbers under the lines at the bottom indicate the time slots. The offline scheme sends $d$ symbols in $\x[b-\delay_L]$}}
        \label{fig:conv2Scheme2}
\end{figure}

We will present an offline coding scheme for the two \messageSizeSequences with rates
\begin{equation}
\label{eq:ratesConverse2}
R_\len^{(1)} = \frac{b-\tau_L+1}{2b-2\tau_L+1.5}, \quad \quad R_\len^{(2)} = \frac{b-\tau_L+2}{2b-2\tau_L+3}
\end{equation}
on \messageSizeSequence $1$ and $2$, respectively. After presenting the scheme for each \messageSizeSequence, we verify that it satisfies the \losslessDelay and \worstCaseDelay constraints.

\para{\textit{Offline scheme for \messageSizeSequence $1$:}} The first $(b-\tau_L)$ \messagePackets are sent in the corresponding channel packets. The \messagePacket $S[b-\tau_L]$ is divided in half to be evenly transmitted over $X[b-\tau_L]$ and $X[b]$. Each of the next $(b-\tau_L)$ channel packets comprises $d$ parity symbols used to decode (a) the first $(b-\tau_L)$ \messagePackets if the corresponding channel packets are lost and (b) $X[b]$ if $X[b-\tau_L]$ and $X[b]$ are both lost. The summation of $X[b-\tau_L]$ and $X[b]$ is later sent in $X[2b]$ to ensure decoding of $S[b-\tau_L]$ within delay $\tau$. The scheme is detailed below and shown in Figure~\ref{fig:conv2Scheme1}:
\begin{itemize}
    \item $S[0]$ and $S[b-\tau_L]$ are each evenly divided into two components of $d/2$ symbols each: $S[0] = (S^{(0)}[0],S^{(1)}[0])$  and $S[b-\tau_L] = (S^{(0)}[b-\tau_L],S^{(1)}[b-\tau_L])$.
    \item For $i \in [b-\tau_L-1]$, $X[i] = S[i]$.
    \item $X[b-\tau_L] = S^{(0)}[b-\tau_L]$.
    \item $X[b] = S^{(1)}[b-\tau_L]$.
    \item $X[b+1] = (S^{(0)}[0],S^{(1)}[0] + S^{(1)}[b-\tau_L])$.
    \item For $i \in \{1,\ldots, b-\tau_L-1\}$, $X[i+b+1] = (X[i+b]+S[i])$.
    \item $X[b-\tau_L+\tau]=X[2b] = ( S^{(0)}[b-\tau_L] + S^{(1)}[b-\tau_L])$.
\end{itemize}

\noindent {\textbf{Decoding:}}
Each \messagePacket is sent within the current and perhaps next $\tau_L$ \channelPackets and is decoded when the transmission is lossless. We now discuss how \messagePackets are recovered within a delay of $\delay$ under lossy conditions. Either $X[0] = S[0]$ is received, or $X[0]$ is lost. In the latter case, both $X[b]=S^{(1)}[b-1]$ and $X[b+1] = (S^{(0)}[0],S^{(1)}[0]+S^{(1)}[b-1])$ are received. Therefore, $S[0]$ is decoded within the delay of $\tau$.
Next, for $i \in \{1,\ldots,b-\tau_L-1\}$, either $X[i] = S[i]$ is received, or both $X[i+b]$ and $X[i+b+1] = (X[i+b]+S[i])$ are received. Thus, $S[i]$ is recovered within delay $(b+1 \le \tau)$.
Either $X[b-\tau_L]=S^{(0)}[b-\tau_L]$ is received, or $X[2b-\tau_L] = \left((S^{(0)}[0],S^{(1)}[0] + S^{(1)}[b-\tau_L]) + \sum_{i=1}^{b-\tau_L-1} S[i]\right)$ is received.
In the latter case, $S[0],\ldots,S[b-\tau_L-1]$ are decoded by time slot $(2b-1)$ and combined with $X[2b-\tau_L]$ to decode $S^{(1)}[b-\tau_L]$. $S^{(1)}[b-\tau_L]$ is then combined with $X[2b]= ( S^{(0)}[b-\tau_L] + S^{(1)}[b-\tau_L])$ to recover $S^{(0)}[b-\tau_L]$ within a delay of $\tau$.
Therefore, $S^{(0)}[b-\tau_L]$ is decoded within delay $\tau$.
Either $X[b]=S^{(1)}[b-\tau_L]$ is received, or $X[2b] = (S^{(0)}[b-\tau_L]+S^{(1)}[b-\tau_L])$ is received. Recall that $S^{(0)}[b-\tau_L]$ is decoded by time slot $2b$. Thus, $S^{(1)}[b-\tau_L]$ is recovered within delay $\tau$.

\para{\textit{Offline scheme for \messageSizeSequence $2$:}} Each \messagePacket $S[i]$ is transmitted in the corresponding channel packet $X[i]$. The next $\tau_L$ channel packets each comprise $d$ parity symbols. These $d\tau_L$ symbols are used to decode (a) the first $(b-\tau_L)$ \messagePackets when the corresponding channel packets are lost, and (b) $S[b]$ when both $X[b-\tau_L] = S[b-\tau_L]$ and $X[b]=S[b]$ are lost.
The sum of $S[b-\tau_L]$ and $S[b]$ is sent in $X[2b]$ to ensure that $S[b-\tau_L]$ is recovered if $X[b-\tau_L]$ is dropped.
The scheme is described in full detail below {and shown in Figure~\ref{fig:conv2Scheme2}}
:
\begin{itemize}
    \item For $i \in [b-\tau_L] \cup \{b\}$, $S[i] = X[i]$.
    \item $X[b+1] = (S[0] + S[b])$.
    \item For $i \in \{1,\ldots,b-\tau_L-1\}$, $X[i+b+1] = (X[b+i] + S[i])$.
    \item $X[2b] = (S[b] + S[b-\tau_L])$.
\end{itemize}

\noindent {\textbf{Decoding:}}
Each \messagePacket is transmitted within the corresponding \channelPacket and is decoded when the transmission is lossless. We now discuss how each \messagePacket is decoded within a delay of $\tau$ under lossy conditions. Either $X[0]=S[0]$ is received or both $X[b]=S[b]$ and $X[b+1] = (S[0] + X[b])$ are received. Consequently, $S[0]$ is decoded within delay $(b+1 \le \tau)$. For $i \in \{1,\ldots,b-\tau_L-1\}$, either $X[i] = S[i]$ is received, or both $X[i+b]$ and $X[i+b+1] = (X[i+b] + S[i])$ is received. Therefore, each $S[i]$ is recovered within delay $(b+1 \le \tau)$. Either $X[b-\tau_L] =S[b-\tau_L]$ is received, or $X[2b-\tau_L]=(S[b] + \sum_{i=0}^{b-\tau_L-1} S[i])$ is received.
In the latter case, $S[0],\ldots,S[b-\tau_L-1]$ are decoded by time slot $(b-\tau_L+\tau)$ and  combined with $X[2b-\tau_L]$ to decode $S[b]$. Then, $S[b]$ and $X[2b] = (S[b]+S[b-\tau_L])$ used to recover $S[b-\tau_L]$. Therefore, $S[b-\tau_L]$ is decoded within delay $\tau$. Either $X[b]=S[b]$ is received, or $X[2b] = (S[b]+S[b-\tau_L])$ is received. In the latter case, subtracting $S[b-\tau_L]$ yields $S[b]$. Hence, $S[b]$ is recovered within a delay of $\tau$.

\para{Proof of the converse result}: The \offop is at least $R_\len^{(1)}$ and $R_\len^{(2)}$ on \messageSizeSequences $1$ and $2$, respectively (i.e., the rate of the offline scheme from Equation~\ref{eq:ratesConverse2}). Next, we present necessary and mutually exclusive conditions on the total number of symbols sent in $X[0],\ldots,X[b-1]$ for a code construction to attain rates at least $R_\len^{(1)}$ and $R_\len^{(2)}$ on the two respective \messageSizeSequences.
The two \messageSizeSequences are the same until time slot $b$.
Therefore, no online coding scheme can satisfy the condition for both \messageSizeSequences.

\para{Condition for rate $R_\len^{(1)}$ on \messageSizeSequence $1$}: Consider an arbitrary coding scheme for \messageSizeSequence $1$. At least $d(b-\tau_L+1)$ symbols are transmitted in $X[b],\ldots,X[\len]$ since $X[0],\ldots,X[b-1]$ could be dropped in a burst. At most, an additional $d(b-\tau_L+.5)$ symbols can be sent over $X[0],\ldots,X[b-1]$ if the rate is at least $R_\len^{(1)}$.

\para{Condition for rate $R_\len^{(2)}$ on \messageSizeSequence $2$}: Consider any coding scheme for \messageSizeSequence $2$. We will show that if
\begin{equation}
    \label{eq:Conditionconv2}
    d' = \sum_{i=0}^{\burst-1} n_i \le d(b-\tau_L+.5)
\end{equation}
then the rate is strictly less than $R_\len^{(2)}$.
At a high level, at least $d(b-\tau_L+2)$ symbols are sent in $X[0],\ldots,X[b-1],X[2b],\ldots,X[\len]$ to satisfy the \worstCaseDelay constraint when $X[b],\ldots,X[2b-1]$ are lost. At least $d(b-\tau_L+1.5)$ symbols must be sent in $X[b],\ldots,X[2b-1]$ for the \losslessDelay and \worstCaseDelay constraints to be satisfied, as will be shown shortly. In total, $d(2b-2\tau_L+3.5)$ symbols are sent, whereas at most $d(2b-2\tau_L+3)$ symbols are transmitted as part of a scheme with a rate of at least $R_\len^{(2)}$.

Next, the fact that sending at most $d(\burst-\tau_L+.5)$ symbols over $\x[0],\ldots,\x[b-1]$ leads to a rate of less than $R_\len^{(2)}$ on \messageSizeSequence $2$ is proven in detail.
Let $\mathcal{S}$ be a random variable drawn uniformly at random from the finite field $\F_q$.
{Recall from Appendix~\ref{sec:prove_thm_osc_opt} that for any $i \in [\len]$, (a) $ H(\s[i]) = H(\mathcal{S}) k_i $, and (b) $H(\x[i]) \le H(\mathcal{S})n_i$ (Equations~\ref{eq:sizeMessage} and~\ref{eq:sizeChannel}).}

{We provide an upper bound on the sizes of the \channelPackets as follows}
\begin{align}
    d(b-\tau_L+2)H\left(\mathcal{S}\right) =&H\left(\s[0:b]\right) \label{eq:recoverMiddleBurst1}\\
    \le& H\left(\s[0:b],\x[0:b-1],\x[2b:b+\delay]\right) \label{eq:recoverMiddleBurst2}\\
    =& H\left(\x[0:b-1],\x[2b:b+\delay]\right) + H\left(\s[0:b]\big |\x[0:b-1],\x[2b:b+\delay]\right) \label{eq:recoverMiddleBurst3}\\
    =& H\left(\x[0:b-1],\x[2b:b+\delay]\right) \label{eq:recoverMiddleBurst4}\\
    \le &H\left(\mathcal{S}\right) \left(\sum_{i = 0}^{b-1} n_i + \sum_{i=2b}^{b+\delay} n_i\right) \label{eq:recoverMiddleBurst5}.
\end{align}
{Equation~\ref{eq:recoverMiddleBurst1} follows from Equation~\ref{eq:sizeMessage}, Equation~\ref{eq:recoverMiddleBurst2} follows from the definition of entropy, Equation~\ref{eq:recoverMiddleBurst3} follows from the chain rule, Equation~\ref{eq:recoverMiddleBurst4} follows from the \worstCaseDelay constraint, and Equation~\ref{eq:recoverMiddleBurst5} follows from Equation~\ref{eq:sizeChannel}.}

Next, we will prove that $H\left(\x[b:2b-1]\right) \ge d(b-\tau_L+1.5)H\left(\mathcal{S}\right)$ as follows
\begin{align}
  H\left(\x[0:b-1],\s[0:b-\delay_L-1]\right) =&H\left(\x[0:b-1]\right) + H\left(\s[0:b-\delay_L-1]\big |\x[0:b-1]\right)\label{eq:BoundChang1}\\
=&H\left(\x[0:b-1]\right) \le d'H\left(\mathcal{S}\right) \label{eq:BoundChang2}\\
H\left(\x[0:b-1],\s[0:b-\delay_L-1]\right) =&H\left(\s[0:b-\delay_L-1]\right) +H\left(\x[0:b-1]\big |\s[0:b-\delay_L-1]\right)\label{eq:BoundChang4} \\
=&d(b-\tau_L)H\left(\mathcal{S}\right) + H\left(\x[0:b-1]\big |\s[0:b-\delay_L-1]\right) \label{eq:BoundChang5}
\end{align}
{where Equation~\ref{eq:BoundChang1} follows from the chain rule, Equation~\ref{eq:BoundChang2} follows from the \losslessDelay constraint and Equation~\ref{eq:Conditionconv2}, Equation~\ref{eq:BoundChang4} follows from the chain rule, and Equation~\ref{eq:BoundChang5} follows from applying Equation~\ref{eq:sizeMessage} to $\s[0],\ldots,\s[b-\delay-1]$.}

Rearranging terms yields
\begin{equation}
\label{eq:boundChanGivMes}
H\left(\x[0:b-1]\big |\s[0:b-\delay_L-1]\right) \le \left(d'-d(b-\tau_L)\right)H\left(\mathcal{S}\right)
\end{equation}

{Next, we bound the sizes of $\x[b],\ldots,\x[2b-1]$ using
\begin{align}
d(b-\tau_L+2)H\left(\mathcal{S}\right) \le & H\left(\s[0:b]\right)
\label{eq:conv2Rear1} \\
 \le &H\left(\s[0:b],\x[0:2b-1]\right)\label{eq:conv2Rear2}\\
 \begin{split}\label{eq:conv2Rear3}
         \le & H\left(\x[b:2b-1]\right) + H\left(\s[0:b-\delay_L-1]\big |\x[b:2b-1]\right) \\
         +&  H\left(\x[0:b-1]\big |\s[0:b-\delay_L-1]\right) + H\left(\s[b-\delay_L:b]\big |\x[0:2b-1]\right)
 \end{split}\\
=& H\left(\x[b:2b-1]\right) + H\left(\x[0:b-1]\big |\s[0:b-\delay_L-1]\right) \label{eq:conv2Rear4}\\
\le &  H\left(\x[b:2b-1]\right) + \left(d'-d(b-\tau_L)\right)H\left(\mathcal{S}\right) \label{eq:conv2Rear5},
\end{align}
where Equation~\ref{eq:conv2Rear1} follows from Equation~\ref{eq:sizeMessage}, Equation~\ref{eq:conv2Rear2} follows from the definition of entropy, Equation~\ref{eq:conv2Rear3} follows from the definition of conditioning, Equation~\ref{eq:conv2Rear4}  follows from the \worstCaseDelay constraint (i.e., $\tau = (\tau_L + \burst)$) and the \losslessDelay (i.e., $\tau_L < \burst$), and Equation~\ref{eq:conv2Rear5} follows from Equation~\ref{eq:boundChanGivMes}.}

Rearranging terms yields
\begin{equation}
    \label{eq:sizeMiddle}
    \begin{aligned}
               &\left(d(2b-2\tau_L+2)-d'\right)H\left(\mathcal{S}\right) \le H\left(\x[b:2b-1]\right) \\
               &\le H\left(\mathcal{S}\right) \sum_{i=b}^{2b-1} n_i.
    \end{aligned}
\end{equation}

The total number of symbols sent in $\x[0:b-1]$ and $\x[2b:b+\delay]$ is at least $d(b-\tau_L+2)$ by Equations~\ref{eq:recoverMiddleBurst1} through \ref{eq:recoverMiddleBurst5}. At least $\left(d(2b-2\tau_L+2)-d'\right)$ symbols are sent in $\x[b:2b-1]$ by Equation~\ref{eq:sizeMiddle}. In total, at least
\begin{equation*}
    d(3b-3\tau_L+4) -d' \ge \big(d(3b-3\tau_L+4) -d(b-\tau_L+.5)\big)=d(2b-2\tau_L+3.5)
\end{equation*}
symbols are sent.
Thus, the rate is strictly lower than $R_\len^{(2)}$.

\para{Summary}:
Any online scheme with rate at least $R_\len^{(1)}$ on \messageSizeSequence $1$ sends at most $d(b-\tau_L+.5)$ symbols over $X[0],\ldots,X[b-1]$. Consequently, its rate is strictly less than $R_\len^{(2)}$ on \messageSizeSequence $2$.

\subsection{Proof of Theorem~\ref{thm:gap} case $\tau_L < (\tau-\burst)$}
\label{sec:conv3}
Let $d$ be an arbitrary positive even integer.
Consider the following two \messageSizeSequences for which the offline construction will be shown below in Figures~\ref{fig:conv3Scheme1} and~\ref{fig:conv3Scheme2} respectively:
\begin{enumerate}
    \item $k^{(1)}_0 = \ldots = k^{(1)}_{b-1} = d$, and $k^{(1)}_b = \ldots = k^{(1)}_\len = 0$.
    \item $k^{(2)}_0 = \ldots = k^{(2)}_{\tau-\tau_L-2}=d$, $k^{(2)}_{\tau-\tau_L-1}=d(\tau_L+1),$ and $k^{(2)}_{\tau-\tau_L} = \ldots = k^{(2)}_\len = 0$.
\end{enumerate}
Before we present the details of the proof, we point out that a similar proof applies to when the sizes of the \messagePackets are approximately equal to those of the \messageSizeSequences, as long as the deviation is small relative to $d$. In addition, the proof extends to scenarios where one of the two above \messageSizeSequence occurs at some point in the transmission proceeded and followed by $\delay$ \messagePackets whose sizes are much less than $d$.

\begin{figure}
            \centering
\includegraphics[angle=0, height=90pt]{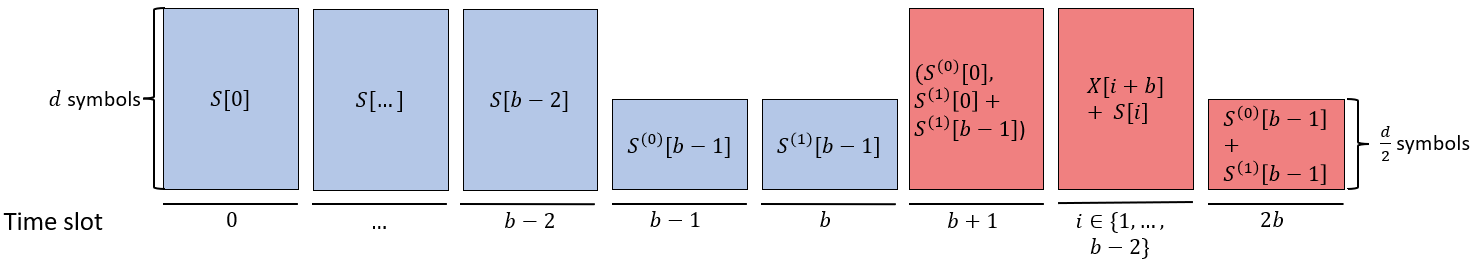}
\caption{{The offline scheme for \messageSizeSequence $1$ for case $\tau_L < (\tau-\burst)$. Blue \channelPackets consist of message symbols, and red \channelPackets consist of parity symbols. The numbers under the lines at the bottom indicate the time slots. The offline scheme sends $\frac{d}{2}$ symbols in $\x[b-1]$. }}
\label{fig:conv3Scheme1}
\end{figure}
\begin{figure}
\centering
\includegraphics[angle=0, height=90pt]{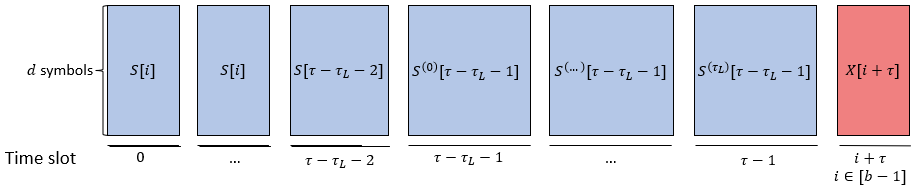}
        \caption{{The offline scheme for \messageSizeSequence $2$ for case $\tau_L < (\tau-\burst)$. Blue \channelPackets consist of message symbols and red \channelPackets consist of parity symbols. The numbers under the lines at the bottom indicate the time slots. The offline scheme sends $d$ symbols in $\x[b-1]$ }}
        \label{fig:conv3Scheme2}
\end{figure}

We will describe an offline coding scheme for \messageSizeSequences $1$ and $2$ with rates
\begin{equation}
\label{eq:ratesConverse3}
R_\len^{(1)} = \frac{b}{2b-.5}, \quad \quad R_\len^{(2)} =  \frac{\tau}{\tau+b}
\end{equation}
on the two respective \messageSizeSequences. We also verify that the \losslessDelay and \worstCaseDelay constraints are satisfied.

\para{\textit{Offline scheme for \messageSizeSequence $1$:}} Each of $S[0],\ldots,S[b-2]$ is transmitted immediately as part of the corresponding channel packet. Then $S[b-1]$ is divided in half and evenly sent over $X[b-1]$ and $X[b]$. The next $(b-1)$ channel packets each comprise $d$ parity symbols. These $d(b-1)$ parity symbols are used to decode (a) \messagePackets $S[0],\ldots,S[b-2]$ when the corresponding channel packets are lost, and (b) $X[b]$ when both $X[b-1]$ and $X[b]$ are lost. The summation of $X[b-1]$ and $X[b]$ is sent in $X[2b]$ to ensure that $S[b-1]$ is decoded within a delay of $\tau$. The scheme is described in detail below {and shown in Figure~\ref{fig:conv3Scheme1}}
:
\begin{itemize}
    \item The \messagePackets $S[0]$ and $S[b-1]$ are divided in half into $S[0] = (S^{(0)}[0],S^{(1)}[0])$ and $S[b-1] = (S^{(0)}[b-1],S^{(1)}[b-1])$ .
    \item For $j \in [b-2]$, $X[j] = S[j]$.
    \item $X[b-1]=S^{(0)}[b-1]$.
    \item $X[b] = S^{(1)}[b-1]$.
    \item $X[b+1] = (S^{(0)}[0],S^{(1)}[0]+S^{(1)}[b-1])$.
    \item For $i \in \{1,\ldots,b-2\}$, $X[i+b+1] = (X[i+b] + S[i])$.
    \item $X[2b] = ( S^{(0)}[b-1]+S^{(1)}[b-1])$.
\end{itemize}

\para{Decoding:}
Each \messagePacket is sent within the current and perhaps next \channelPackets and is decoded when the transmission is lossless.
We now discuss how each \messagePacket is decoded within delay $\delay$ under lossy conditions. Either $X[0]=S[0]$ is received, or both $X[b]=S_1[b-1]$ and $X[b+1] = (S^{(0)}[0],S^{(1)}[0]+S^{(1)}[b-1])$ are received. Thus, $S[0]$ is decoded within a delay of $(b+1 \le \tau)$.
For $j \in \{1,\ldots,b-2\}$, either $X[j]=S[j]$ is received, or both $X[j+b]$ and $X[j+b+1] = (X[j+b] + S[j])$ are received. Therefore, $S[j]$ is decoded within delay $(b+1 \le \tau)$. Either $X[b-1]=S^{(0)}[b-1]$ is received, or $X[2b-1]$ is received. In the latter case, $S[0],\ldots,S[b-2]$ are decoded by time slot $(2b-1)$ and are combined with $X[2b-1]= \left((S^{(0)}[0],S^{(1)}[0]+S^{(1)}[b-1]) + \sum_{i=1}^{b-2} S[i]\right)$ to recover $S^{(1)}[b-1]$. The receiver then decodes $S^{(0)}[b-1] = (X[2b]-S^{(1)}[b-1])$ within delay $(b+1 \le \tau)$. Either $X[b] = S^{(1)}[b-1]$ is received, or $X[2b] =(S^{(0)}[b-1]+S^{(1)}[b-1])$ is received and combined with $S^{(0)}[\burst-1]$ to recover $S^{(1)}[b-1]$ within delay $\tau$.

\para{\textit{Offline scheme for \messageSizeSequence $2$:}} Each of $S[0],\ldots,S[\tau-\tau_L-2]$ is transmitted within the corresponding channel packet. The symbols of $S[\tau-\tau_L-1]$ are evenly divided into $(\tau_L+1)$ components sent over $X[\tau-\tau_L-1],\ldots,X[\tau-1]$ respectively. Each of $X[\tau],\ldots,X[\tau+b-1]$ comprises $d$ symbols, which creates $d$ blocks of the $[\tau+b,\tau]$ systematic block codes (described in Section~\ref{sec:conv1}). The scheme is presented in detail below {and shown in Figure~\ref{fig:conv3Scheme2}}
:
\begin{itemize}
    \item For $j \in [\tau-\tau_L-2]$, $X[j] = S[j]$.
    \item The \messagePacket $S[\tau-\tau_L-1]$ is evenly divided into $(\tau_L+1)$ components of size $d$: $(S^{(0)}[\tau-\tau_L-1],\ldots,S^{(\tau_L)}[\tau-\tau_L-1])$.
    \item For $j \in \{\tau-\tau_L-1,\ldots,\tau-1\}$, $X[j] = S^{(j-\tau+\tau_L+1)}[\tau-\tau_L-1]$.
    \item For each $z \in [d-1]$, an instance of the block code from \cite{krishnan2019low} is created wherein $(X_z[0],\ldots,X_z[\tau-1])$ is mapped to $(X_z[0],\ldots,X_z[\tau-1],p_0^{(z)},\ldots,p_{b-1}^{(z)})$.
    \item For $j \in [b-1]$, $X[\tau+j] = (p_j^{(0)},\ldots,p_j^{(d-1)})$.
\end{itemize}

\para{Decoding:}
Each \messagePacket is sent over the current and perhaps next $\delay_L$ \channelPackets and is decoded when the transmission is lossless.
Under lossy conditions, the block code $(X_z[0],\ldots,X_z[\tau-1],p_0^{(z)},\ldots,p_{b-1}^{(z)})$ is used for decoding. For $z \in [d-1]$:
(a) Each symbol $X_z[i]$, for $i \in [b-1]$, is decoded within a delay of $\tau$.
(b) Each symbol $X_z[i]$, for $i \in [\tau-1] \setminus [b-1]$, is decoded by time slot $(\tau+b-1)$. Thus, the \worstCaseDelay constraint is satisfied.

\para{Proof of the converse result}: The rates $R_\len^{(1)}$ and $R_\len^{(2)}$ of the above construction (Equation~\ref{eq:ratesConverse3}) for \messageSizeSequences $1$ and $2$, respectively, serve as a lower bound on the \offop for the two \messageSizeSequences. Next, we present mutually exclusive conditions on the number of symbols transmitted in the first $b$ channel packets to have rates at least $R_\len^{(1)}$ or $R_\len^{(2)}$ on \messageSizeSequences $1$ or $2$, respectively. The online coding schemes cannot differentiate between the two \messageSizeSequences before the time slot $b$. Hence, the number of symbols sent in $X[0],\ldots,X[b-1]$ by any online scheme violates the condition for at least one \messageSizeSequence.

\para{Condition for rate $R_\len^{(1)}$ on \messageSizeSequence $1$}: Consider any coding scheme for \messageSizeSequence $1$. At least $db$ symbols are transmitted in $X[b],\ldots,X[\len]$ in case there is a burst loss of $X[0],\ldots,X[b-1]$. The rate is at least $R_\len^{(1)}$, so at most $d(b-.5)$ additional symbols are sent in $X[0],\ldots,X[b-1]$.

\para{Condition for rate $R_\len^{(2)}$ on \messageSizeSequence $2$}: Consider any coding scheme for \messageSizeSequence $2$.
We will demonstrate that if
\begin{equation}
\label{eq:conv3Cond1}
\sum_{i=0}^{\burst-1} n_i \le d(b-.5)
\end{equation}
then the rate is strictly less than $R_\len^{(2)} = \frac{\tau}{\tau+b}$ { in two steps. First, we will show that all symbols are transmitted by $X[\tau+b-1]$ without loss of generality. Second, we prove that strictly more than $db$ symbols may be lost.
At least $d\tau$ additional symbols are sent to meet the \worstCaseDelay constraint, leading to a lower rate than $R_\len^{(2)}$.}

\para{Step 1:} If $\x[\delay+\burst-1]$ is lost, then $\x[0:\delay-1]$ are received, which yields $\s[0:\tau-\tau_L-1]$ by the \losslessDelay constraint.
Thus, all symbols sent after the time slot $(\delay+\burst)$ can instead be sent in $\x[\delay+\burst-1]$.

\para{Step 2:} Consider the following erasure channels $C_{i}$ for $i \in [\tau+b-1]$. Each $C_{i}$ introduces bursts of packet losses in $\{X[j],\ldots,X[j+b-1] \mid j \equiv i \mod (\tau+b)\}$ and results in $l_i$ lost (dropped) symbols.\footnote{A similar argument was used to show the upper bound on rate of $\frac{\tau}{\tau+b}$ in \cite{martinian2007delay}.}
At least $d(\tau+b)$ symbols are sent in total due to the upper bound on the rate of $\frac{\tau}{\tau+b}$, leading to
\begin{align}
\sum_{i=0}^{\tau+b-1} l_i & \ge db(\tau+b) \label{eq:per}\\
\sum_{i=1}^{\tau+b-1} l_i & \ge db(\tau+b-1) +.5d \label{eq:perLestFirst}\\
\frac{1}{\tau+b-1}\sum_{i=1}^{\tau+b-1} l_i & \ge db +\frac{.5d}{\tau+b-1},
\end{align}
{where Equation~\ref{eq:per} follows from each packet (and hence each symbol) being dropped by exactly $b$ channels, and Equation~\ref{eq:perLestFirst} follows from Equation~\ref{eq:conv3Cond1}.}

Hence, there is some $i \in \{1,\ldots,\tau+b-1\}$ for which
$l_i \ge (db + \frac{.5d}{\tau+b-1}).$
In order to satisfy the \worstCaseDelay constraint over channel $C_{i}$, at least $d\tau$ symbols are received outside of the channel packets dropped by $C_{i}$.
Thus, the total number of symbols sent is at least $d(\tau+b+\frac{.5}{\tau+b-1})$. In contrast, at most $d(\tau+b)$ symbols are sent if the rate is at least $R_\len^{(2)}$.

\para{Summary}: Any online coding scheme with a rate of at least $R_\len^{(1)}$ on \messageSizeSequence $1$ sends at most $d(b-.5)$ symbols in $X[0],\ldots,X[b-1]$. Consequently, its rate is strictly lower than $R_\len^{(2)}$ on \messageSizeSequence $2$.

\section*{Acknowledgment}
This work was funded in part by an NSF grant (CCF-1910813). The authors thank Francisco Maturana for his feedback in editing this work.



\bibliographystyle{IEEEtran}
\bibliography{IEEEabrv,main}

\mrem{\noindent \textbf{Michael Rudow} is a final-year Ph.D. student in the Computer Science Department at Carnegie Mellon University, advised by Professor Rashmi Vinayak. Before starting at Carnegie Mellon, he earned his bachelor’s degree in computer science and master’s degree in mathematics at the University of Pennsylvania. He is broadly interested in the intersection of coding theory and machine learning.}

\mrem{$\\~\\$}

\mrem{\noindent \textbf{K.V. Rashmi} is an assistant professor in the Computer Science department at Carnegie Mellon University. Rashmi is a recipient of Sloan Research Fellowship 2023, Meta Research Award 2022, VMWare Systems Research Award 2021, NSF CAREER Award 2020-25, Tata Institute of Fundamental Research Memorial Lecture Award 2020, Facebook Distributed Systems Research Award 2019, Google Faculty Research Award 2018, and Facebook Communications and Networking Research Award 2017. Her PhD thesis was awarded the UC Berkeley Eli Jury Dissertation Award 2016, and her work has received USENIX NSDI 2021 Community (Best Paper) Award, and IEEE Data Storage Best Paper and Best Student Paper Awards for the years 2011/2012. Rashmi received her Ph.D. from UC Berkeley in 2016, and was a postdoctoral scholar at UC Berkeley during 2016-17. During her Ph.D. studies, Rashmi was a recipient of Facebook Fellowship 2012-13, the Microsoft Research PhD Fellowship 2013-15, and the Google Anita Borg Memorial Scholarship 2015-16. Her research interests broadly lie in information/coding theory and computer/networked systems.}










\end{document}